\documentclass{emulateapj}
\usepackage{natbib}
\usepackage{lscape}
\usepackage{longtable}
\newcommand {\aplt} {\ {\raise-.5ex\hbox{$\buildrel<\over\sim$}}\ } 
\newcommand {\apgt} {\ {\raise-.5ex\hbox{$\buildrel>\over\sim$}}\ } 

\def\sun{\hbox{$\odot$}}

\newcommand{\gps}{\ensuremath{g_{\rm P1}}}
\newcommand{\rps}{\ensuremath{r_{\rm P1}}}
\newcommand{\ips}{\ensuremath{i_{\rm P1}}}
\newcommand{\zps}{\ensuremath{z_{\rm P1}}}
\newcommand{\yps}{\ensuremath{y_{\rm P1}}}

\newcommand{\PS}{\protect \hbox {Pan-STARRS1}}

\newcommand{\qso}{358~}
\newcommand{\rr}{37~}
\newcommand{\mm}{53~}
\newcommand{\agn}{305~}

\newcommand{\ms}{17~}

\slugcomment{Accepted for publication in ApJ}
\shorttitle{\textsl{GALEX} Time Domain Survey}
\shortauthors{Gezari et al.}

\begin{document}

\title{The \textsl{GALEX} Time Domain Survey I. Selection and Classification of Over a Thousand UV Variable Sources}

\author{S. Gezari\altaffilmark{1}, D.C. Martin\altaffilmark{2}, K. Forster\altaffilmark{2}, J.D. Neill\altaffilmark{2}, M. Huber\altaffilmark{3}, T. Heckman\altaffilmark{4}, L. Bianchi\altaffilmark{4}, P. Morrissey\altaffilmark{2}, S.G. Neff\altaffilmark{5}, M. Seibert\altaffilmark{6}, D. Schiminovich\altaffilmark{7}, Wyder, T.K.\altaffilmark{2}, W. S. Burgett,\altaffilmark{3}
K. C. Chambers,\altaffilmark{3} 
N. Kaiser,\altaffilmark{3}
E. A. Magnier,\altaffilmark{3}
P. A. Price,\altaffilmark{8} and
J.L. Tonry\altaffilmark{3} 
}
\altaffiltext{1}{Department of Astronomy, University of Maryland, College Park, MD 20742-2421, USA \email{suvi@astro.umd.edu}}
\altaffiltext{2}{California Institute of Technology, 
MC 249-17, 1200 East California Boulevard, Pasadena, CA  91125, USA}
\altaffiltext{3}{Institute for Astronomy, University of Hawaii, 2680 Woodlawn Drive, Honolulu HI  96822, USA}
\altaffiltext{4}{Department of Physics and Astronomy,
        Johns Hopkins University,
        3400 North Charles Street,
        Baltimore, MD 21218, USA}
\altaffiltext{5}{Laboratory for Astronomy and Solar Physics, NASA Goddard Space Flight Center, Greenbelt, MD 20771, USA}
\altaffiltext{6}{Observatoires of the Carnegie Institute of Washington, Pasadena, CA 90095, USA}
\altaffiltext{7}{Department of Astronomy, Columbia University, New York, NY 10027, USA}

\altaffiltext{8}{Department of Astrophysical Sciences, Princeton University, Princeton, NJ 08544, USA}

\begin{abstract}
We present the selection and classification of over a thousand ultraviolet (UV) variable sources discovered in $\sim$ 40 deg$^{2}$ of \textsl{GALEX} Time Domain Survey (TDS) $NUV$ images observed with a cadence of 2 days and a baseline of observations of $\sim 3$ years.  The \textsl{GALEX} TDS fields were designed to be in spatial and temporal coordination with the Pan-STARRS1 Medium Deep Survey, which provides deep optical imaging and simultaneous optical transient detections via image differencing.  We characterize the \textsl{GALEX} photometric errors empirically as a function of mean magnitude, and select sources that vary at the 5$\sigma$ level in at least one epoch.    We measure the statistical properties of the UV variability, including the structure function on timescales of days and years.  We report classifications for the \textsl{GALEX} TDS sample using a combination of optical host colors and morphology, UV light curve characteristics, and matches to archival X-ray, and spectroscopy catalogs.  We classify 62\% of the sources as active galaxies (\qso quasars and \agn active galactic nuclei), and 10\% as variable stars (including \rr RR Lyrae, \mm M dwarf flare stars, and 2 cataclysmic variables).  We detect a large-amplitude tail in the UV variability distribution for M-dwarf flare stars and RR Lyrae, reaching up to $|\Delta m| = 4.6$ mag and 2.9 mag, respectively.  The mean amplitude of the structure function for quasars on year timescales is 5 times larger than observed at optical wavelengths.  The remaining unclassified sources include UV-bright extragalactic transients, two of which have been spectroscopically confirmed to be a young core-collapse supernova and a flare from the tidal disruption of a star by dormant supermassive black hole.  We calculate a surface density for variable sources in the UV with $NUV < 23$ mag and $|\Delta m| > 0.2$ mag of $\sim$ 8.0, 7.7, and 1.8 deg$^{-2}$ for quasars, AGNs, and RR Lyrae stars, respectively.  We also calculate a surface density rate in the UV for transient sources, using the effective survey time at the cadence appropriate to each class, of $\sim 15$ and $52$ deg$^{-2}$ yr$^{-1}$ for M dwarfs and extragalactic transients, respectively.
\end{abstract}


\section{Introduction}

Unlike the optical, X-ray, and $\gamma$-ray sky, which have been systematically studied in the time domain in the search for supernovae (SNe) and gamma-ray bursts (GRBs), the wide-field UV time domain is a relatively unexplored parameter space.  The launch of the \textsl{GALEX} satellite with its 1.25 deg diameter field of view, and limiting sensitivity per 1.5 ks visit of $\sim$23 mag in the $FUV$ ($\lambda_{\rm eff}=1539$ \AA) and $NUV$ ($\lambda_{\rm eff}=2316$ \AA) \citep{Martin2005, Morrissey2007}, enabled the discovery of UV variable sources in repeated observations over hundreds of square degrees for the first time.  

The UV waveband is particularly sensitive to hot ($\approx 10^{4}$ K) thermal emission from such transient and variable phenomena as young core-collapse SNe, the inner regions of the accretion flow around accreting supermassive black holes (SMBHs), and the flaring states of variable stars.  The characterstic timescales of variables and transients in the UV range from minutes to years.  M dwarf flare stars have strong magnetic activity that manifests itself in flares of thermal UV emission on the timescale of minutes \citep{Kowalski2009}.  RR Lyrae stars have periodic pulsations which drive temperature fluctuations from $\sim 6000$ to 8000 K that cause periodic variability in the UV on a timescale of $\sim 0.5$ d \citep{Wheatley2012}.  Core-collapse supernovae (SNe) remain bright in the UV for hours up to several days following shock breakout, depending on the radius of the progenitor star \citep{Nakar2010, Rabinak2011}, and the presence of a dense wind \citep{Ofek2010, Chevalier2011, Svirski2012}.  Active galactic nuclei (AGN) and quasars demonstrate stronger variability with decreasing wavelength and longer timescales of years \citep{VandenBerk2004}.

UV variability studies of \textsl{GALEX} data observed as part of the All-Sky, Medium, and Deep Imaging baseline mission surveys (AIS, MIS, DIS) from 2003 to 2007, yielded the detection of M-dwarf flare stars \citep{Welsh2007}, RR Lyrae stars, AGN, and quasars \citep{Welsh2005, Wheatley2008, Welsh2011}, and flares from the tidal disruption of stars around dormant supermassive black holes \citep{Gezari2006, Gezari2008a, Gezari2009}.  Serendipitous overlap of 4 \textsl{GALEX} DIS fields with the optical CFHT Supernova Legacy Survey, enabled the extraction of simultaneous optical light curves from image differencing for 2 of the tidal disruption event (TDE) candidates \citep{Gezari2008a}, and enabled the association of transient UV emission with two Type IIP supernovae (SNe) within hours of shock breakout \citep{Schawinski2008, Gezari2008b}.  Chance overlap of \textsl{GALEX} observations with the survey area of the optical Palomar Transient Factory (PTF) detected a Type IIn SN whose rising UV emission over a few days was interpreted as a delayed shock breakout through a dense circumstellar medium \citep{Ofek2010}. 

Motivated by the promising results from the analysis of random repeated \textsl{GALEX} observations, and the demonstrated value of overlap with optical time domain surveys, we initiated a dedicated \textsl{GALEX} Time Domain Survey (TDS) to systematically study UV variability on timescales of days to years with multiple epochs of $NUV$ images observed with a regular cadence of 2 days.  The \textsl{GALEX} TDS fields were selected to overlap with the Pan-STARRS1 Medium Deep Survey (PS1 MDS) \citep{Kaiser2010}.  \textsl{GALEX} TDS and PS1 MDS are well-matched in field of view, sensitivity, and cadence (shown in Table \ref{tdsmds}).  Here we present the analysis of 42 \textsl{GALEX} TDS fields which intersect with the PS1 MDS footprint, for a total area on the sky of 39.91 deg$^{2}$, which were monitored over a baseline of 3.32 years (February 2008 $-$ June 2011).
In this paper we use PS1 MDS deep stack catalogs to characterize the optical hosts of \textsl{GALEX} TDS sources.   Simultaneous UV and optical variability of \textsl{GALEX} TDS sources culled from matches with the PS1 transient alerts \citep{Huber2011} will be presented in future papers.

The paper is structured as follows.  In \S\ref{sec:galex} we describe the \textsl{GALEX} TDS survey design, and in \S\ref{sec:stats} we describe our statistical methods for selecting UV-variable sources and characterizing their UV variability.  In \S\ref{sec:opt} we describe the multiwavelength catalog data used to identify the hosts of the UV-variable sources, including archival optical data, a deep PS1 MDS catalog, and archival redshift and X-ray catalogs.  In \S\ref{sec:class} we describe our sequence of steps for classifying the \textsl{GALEX} TDS sources, in \S\ref{sec:disc} we summarize our classification results, and the UV variability properties of our classified sources.   In \S\ref{sec:conc} we conclude with implications for future surveys.

\section{\textsl{GALEX} TDS Observations} \label{sec:galex}

\begin{figure}
\plotone{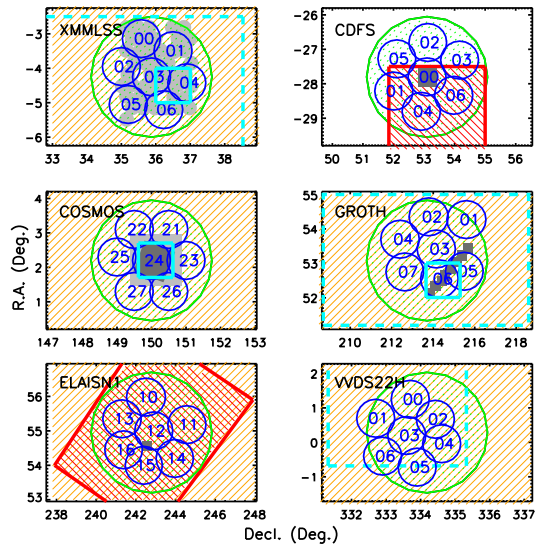}
\caption{\textsl{GALEX} TDS $1.1\deg$ diameter field pointings shown in blue, and the PS1 MDS $3.5\deg$ diameter field pointings shown in green.  Orange hatched regions show the coverage of SDSS in the optical, red hatched regions shows the coverage of SWIRE in the optical, and cyan rectangles indicate the coverage of the CFHTLS Deep (solid lines) and Wide (dashed lines) surveys in the optical.  Light grey regions show the coverage of XMM-Newton X-ray observations, and dark grey regions show the coverage of Chandra X-ray observations.
\label{fig:fields}
}
\end{figure}

\textsl{GALEX} TDS monitored 6 out of 10 total PS1 MDS fields, with 7 \textsl{GALEX} TDS pointings (labeled PS\_{\it fieldname}\_MOS{\it pointing}) at a time to cover the PS1 7 deg$^{2}$ field of view.  During the window of observing visibility of each \textsl{GALEX} TDS field (from $2-4$ weeks, $1-2$ times per year), they were observed with a cadence of 2 days, and a typical exposure time per epoch of 1.5 ks (or a 5$\sigma$ point-source limit of $m_{AB}\sim 23.3$ mag), with a range from 1.0 to 1.7 ks.  The $NUV$ detector developed a problem on 2010 May 4 during observations of PS\_ELAISN1, and so we do not include epochs observed between this time and when the the instrument was fixed on 2010 June 23 in our analysis.  Figure \ref{fig:fields} shows the position of the \textsl{GALEX} TDS fields relative to the PS1 MDS fields, and Table \ref{tab1} lists the R.A. and Dec of their centers, the Galactic extinction ($E(B-V)$) for each field from the \citet{Schlegel1998} dust maps, and the number of epochs per field.  The median number of epochs per field is 24.   PS\_CDFS\_MOS00 is an exception with 114 epochs, because it was monitored with a rapid cadence ($|\Delta t| \sim 3$ hours) over a period of 10 days in 2010 November.  Some offsets in the \textsl{GALEX} pointings from the footprint of the PS1 MDS fields were necessary in order to avoid UV-bright stars in the field of view that would violate the detector's bright-source count limits.  

\begin{figure}
\plotone{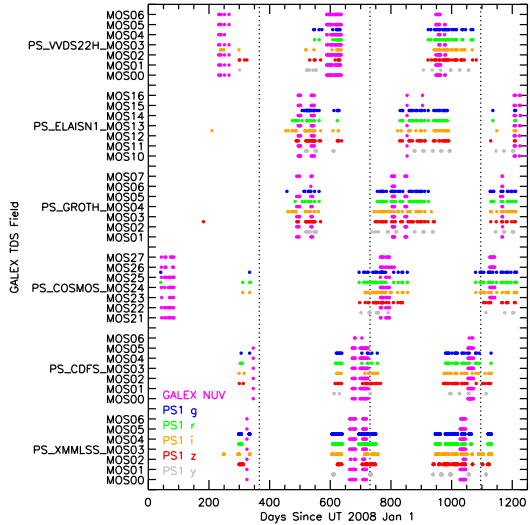}
\caption{Dates of \textsl{GALEX} TDS $NUV$ observations compared to PS1 MDS observations in the \gps, \rps, \ips, \zps, and \yps\ bands.  Dotted lines show yearly intervals.
\label{fig:dates}
}
\end{figure}

Figure \ref{fig:dates} shows the temporal sampling of \textsl{GALEX} TDS observations in the $NUV$ in comparison to the PS1 MDS observations in the \gps, \rps, \ips, \zps, and \yps\ bands from February 2008 to June 2011.  PS1 began taking commissioning data of the MDS fields in May 2009, but did not begin full survey operations until a year later.  The \textsl{GALEX} $FUV$ detector became non-operational in May 2009, and so we only include $NUV$ images in our study.

\section{Statistical Measurements} \label{sec:stats}

\subsection{Selection of Variable Sources}
Since most galaxies are unresolved by the \textsl{GALEX} $NUV$ 5.3 arcsec full-width at half-maximum (FWHM) point spread function (PSF), we can use simple aperture photometry instead of image differencing to measure variability.  We create a master list of unique source positions from the pipeline-generated catalogs \citep{Morrissey2007} for all the individual epochs, as well as deep stacks of all the epochs, using a clustering radius of 5 arcsec.  This radius is chosen such that for the typical astrometry error of \textsl{GALEX} of $0\farcs5$, the Bayesian probability that the match is real is larger than the Bayesian probability that the match is spurious \citep{Budavari2008}. The final master list includes 419,152 sources.

In order to select intrinsically variable sources in our survey, we first need to characterize the photometric errors.  Although the \textsl{GALEX} images are Poisson-limited, the Poisson error underestimates the total error in the \textsl{GALEX} catalog magnitudes by a factor of $\sim 2$ \citep{Trammell2007}.  This discrepancy is attributed to systematic errors such as uncertainties in the detector background and flat-field.  
Thus, we measure the photometric error {\it empirically} by calculating the standard deviation of aperture magnitudes in bins of mean magnitude, $\langle m \rangle$.  We only include objects in the pipeline-generated catalogs that are detected in all or $\ge$10 epochs.  In each bin of $N$ objects  with $\langle m \rangle_{i}=(\sum_{k=1}^{n} m_{i,k})/n$ (each bin typically has $N$ = 50 to 1000 sources), we calculate for each epoch $k$ of a total of $n$ epochs, 

\begin{equation}
\sigma(\langle m \rangle,k) = \sqrt{\frac{1}{N-1} \sum_{i=1}^{N} (m_{i,k}-\langle m \rangle_{i})^{2}}
\end{equation}

\noindent where $m_{i,k}$ is the magnitude (in the AB system) given by $m_{i,k} = -2.5 \log (f_{6}) + zp + C_{\rm ap} - 8.2 E(B-V)$, $f_{6}$ is the background-subtracted flux in a 6 arcsec radius aperture, $zp$ = 20.08, the aperture correction is $C_{\rm ap} = -0.23$ mag \citep{Morrissey2007}, and we correct for Galactic extinction using the values for $E(B-V)$ listed in Table \ref{tab1}.   
We use 3$\sigma$ clipping to remove outliers in the calculation of $\sigma(\langle m \rangle,k)$ which can arise from artifacts.  

\begin{figure}
\plotone{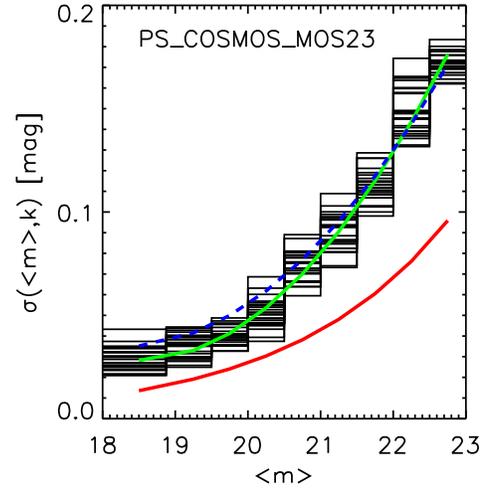}
\caption{Empirical determination of 1$\sigma$ photometric errors as a function of mean magnitude from the standard deviation of sources detected by the pipeline catalogs in $\ge$ 10 epochs for one of the \textsl{GALEX} TDS fields PS\_COSMOS\_MOS23.  Solid green line shows a quadratic fit to the error function for the field PS\_COSMOS\_MOS23, and dashed blue line shows a quadratic fit to the median error function for all of the \textsl{GALEX} TDS fields.  Solid red line shows the expected 1$\sigma$ Poisson error, which underestimates the total photometric error by a factor of $\sim 2$.
\label{fig:sig}
}
\end{figure}

\begin{figure}
\plotone{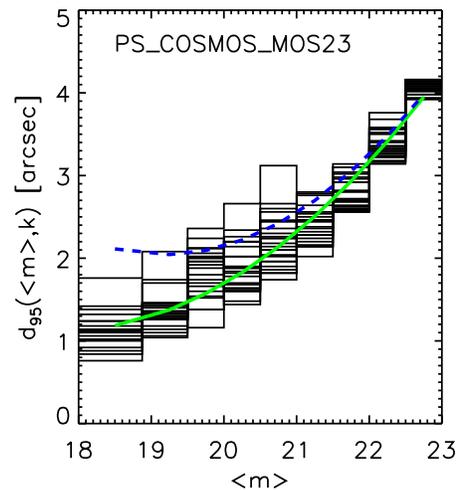}
\caption{Maximum spatial separation from the mean for 95\% of the sources as a function of mean magnitude from the cumulative distribution of sources detected by the pipeline catalogs in $\ge$ 10 epochs for one of the \textsl{GALEX} TDS fields PS\_COSMOS\_MOS23.  Solid green line shows a quadratic fit to the error function for the field PS\_COSMOS\_MOS23, and dashed blue line shows a quadratic fit to the median distance function for all of the \textsl{GALEX} TDS fields.  Note that due to systematic differences in the PSF between fields, the distance function radius for PS\_COSMOS\_MOS23 is up to $\sim 1$ arcsec smaller than the median for all fields, since for $\langle m \rangle \aplt 20$ mag, $d_{95} \sim 2$ arcsec in the PS\_XMMLSS, PS\_CDFS, PS\_GROTH, and PS\_ELAISN1 fields and $d_{95} \sim 3$ arcsec in the PS\_VVDS22H fields.
\label{fig:dist}
}
\end{figure}

The astrometric precision depends on the signal-to-noise of the source, thus we also {\it empirically} measure a magnitude-dependent clustering radius.  We do so by measuring the cumulative distribution of spatial separations between the position in each epoch and the mean position for sources in bins of $\langle m \rangle$, and record $d_{95}(\langle m \rangle,k)$, the value for which 95\% of sources have a separation less than or equal to that value.   The resulting value for $d_{95}$ is a strong function magnitude, increasing from $\sim$ 1 arcsec for $\langle m \rangle = 18.0$ mag to $\sim$ 4 arcsec for $\langle m \rangle = 23.0$ mag.  Figures \ref{fig:sig} and \ref{fig:dist} show $\sigma(\langle m \rangle,k)$ and $d_{95}(\langle m \rangle,k)$ for an example \textsl{GALEX} TDS field PS\_COSMOS\_MOS23, a quadratic fit to the median function for all epochs in that field, and the median function fit over all fields.

In our master list of source positions, we include all sources detected, including sources detected in only one epoch, and fix the centroid to the epoch for which the source is detected with maximum flux.  We measure forced aperture magnitudes at the positions of each source in epochs where the source was not detected by the pipeline or the spatial separation of the matched source is greater than $d_{95}(\langle m \rangle,k)$.    When the aperture magnitude is fainter than $m_{\rm lim}$ in an epoch, it is flagged as an upper limit and replaced with $m_{\rm lim}$, where $m_{\rm lim} = -2.5 \log (5 \sqrt{(B_{\rm sky}N_{\rm pix}/T_{\rm exp})} + zp + C_{\rm ap}$, where $B_{\rm sky}=3\times10^{-3}$ counts s$^{-1}$ pixel$^{-1}$, $N_{\rm pix} = 16\pi$ and $T_{\rm exp}$ is the exposure time of that epoch in seconds.  

We select sources that have at least one epoch for which $|m_{k} - \langle m \rangle| > 5\sigma(\langle m \rangle,k)$, where $\langle m \rangle$ is calculated only from epochs that have a magnitude above the detection limit of that epoch.   We use this selection method to be sensitive to short-term and long-term variability, as well as transients.  This 5$\sigma$ selection is quite conservative, and requires variability amplitudes increasing from $|\Delta m|> 0.1$ mag for $\langle m \rangle \sim 18$ mag up to $|\Delta m| > 1.0$ mag for $\langle m \rangle \sim 23$ mag.  

We make the following cuts to the $5\sigma$ variable source sample to remove artifacts:
\newcounter{bean}
\begin{list}{\roman{bean})}{\usecounter{bean}\setlength{\rightmargin}{\leftmargin}}
\item We remove sources with pipeline artifact flags indicating window bevel reflections or ghosts from the dichroic beam splitter. 
\item We remove the brightest objects, with $\langle m \rangle < 18.0$, due to the large area subtended by the PSF which causes uncertainty in the background subtraction.
\item We select sources within a radius of $0.55$ deg of the center of the field, in order to avoid glints and PSF distortions, which are more prominent on the edges of the image, from un-corrected spatial distortions of photons recorded by the detectors.  
\item  We do not include objects that are within 1.5 arcmin of a $m < 17$ mag source, to avoid regions affected by the bright source's PSF and ghost artifacts.  Ghost artifacts can appear within $30-60$ arcsec above and below a bright source in the Y detector direction.  Ghosts are point-like, and thus can only be identified from their Y detector position relative to a bright source.  While ghosts do not usually appear in the \textsl{GALEX} pipeline catalogs, we apply this cut since our forced aperture photometry could mistake ghosts for transient sources.  
\item We veto objects for which in the epoch of maximum $|m_{k} - \langle m \rangle|/\sigma(\langle m \rangle,k)$ or maximum flux, the aperture flux ratio of the object has $f_{6}/f_{3.8} > R_{90}$, where $f_{6}$ is the 6.0 arcsec radius aperture flux, $f_{3.8}$ is the 3.8 arcsec radius aperture flux, and $R_{90}$ is the maximum cumulative aperture flux ratio measured for 90\% of the sources in the reference source sample used to calculate $\sigma(\langle m \rangle,k)$ in that epoch.  This cut removes fluctuations in the background due to reflections from bright stars just outside the field-of-view, as well as epochs where the PSF is distorted due to a degradation in resolution which sometimes occurs in the Y detector direction.  This also vetoes cases when the pipeline shreds a source into multiple sources, and the source is detected as variable because the center of the aperture is off-center from the peak source flux.
\item Finally, we visually inspect all of the remaining variable sources to remove any remaining artifacts that passed through the cuts above.  
\end{list}

\begin{figure}
\plotone{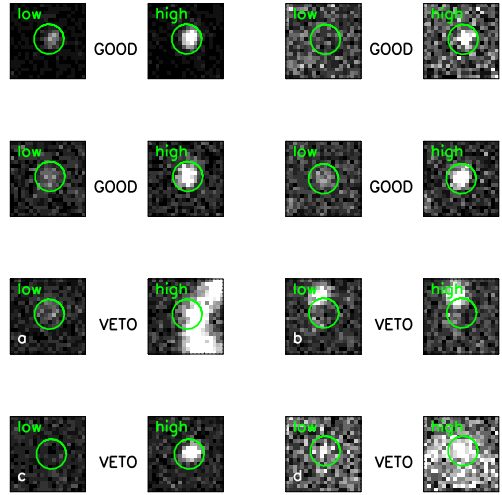}
\caption{{\it Left}:  Gallery of good and vetoed variable sources during their epochs of minimum (``low'') and maximum (``high'') flux.  Green circle shows a 6 arcsec (4 pixel) aperture radius.  Sources vetoed for $f_{6}/f_{3.8} > R_{90}$ shown in panels $a$ and $b$, a source vetoed as a likely ghost within 1.5 arcmin of a $m < 17$ mag source shown in panel $c$, and a manually vetoed source shown in panel $d$.  For each source the grayscale is linear, and is scaled to the peak of the source in its ``high'' state. 
\label{gallery}
}
\end{figure}

Figure \ref{gallery} shows a gallery of good sources and vetoed variable sources from our automated cuts (for a bright reflection in panel a, an off-center source in panel b, and a likely ghost in panel c) and manual cuts (for a diffuse reflection in panel d).  Our final \textsl{GALEX} TDS 5$\sigma$ variable sample after the cuts listed above has a total of 1078 sources.

\subsection{Variability Statistics}

\begin{figure}
\plotone{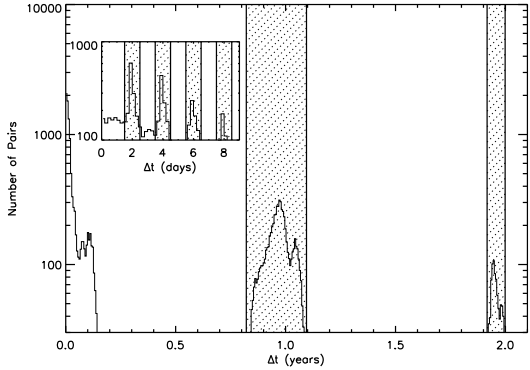}
\caption{Histogram of the time intervals between all pairs of observations for the 42 fields in \textsl{GALEX} TDS.  Hatched regions show the time intervals over which the structure function is measured for all of the sources.
\label{fig:delta}
}
\end{figure}

We characterize the variability of each 5$\sigma$ variable UV source using several statistical measures.  We measure the structure function (following \cite{diClemente1996}),
\begin{equation}
V(\Delta t) = \sqrt{ \frac{\pi}{2}\langle|\Delta m_{ij}|\rangle_{\Delta t}^2 - \langle \sigma_{i}^2+\sigma_{j}^2 \rangle_{\Delta t}}
\end{equation}

\noindent where brackets denote averages for all pairs of points on the light curve of an individual source with $i<j$ and $t_{j}-t_{i} = \Delta t$.  The 2 day cadence of the observations combined with the seasonal visibility of the fields results in a distribution of time intervals between observations (shown in Figure \ref{fig:delta}) that fall into 6 characteristic timescale bins:  $\Delta t_{\rm 2d}$ = $2 \pm 0.5$ d, $\Delta t_{\rm 4d}$ = $4 \pm 0.5$ d, $\Delta t_{\rm 6d}$ = $6 \pm 0.5$ d, $\Delta t_{\rm 8d}$ = $8 \pm 0.5$ d, $\Delta t_{\rm 1yr} = 0.96 \pm 0.14$ yr, and $\Delta t_{\rm 2yr} = 1.96 \pm 0.04$ yr.  We measure the structure function in these 6 bins, and define $S_{\rm d}$ to be the maximum value of the structure function evaluated for $\Delta t_{\rm 2d}, \Delta t_{\rm 4d}, \Delta t_{\rm 6d}$, and $\Delta t_{\rm 8d}$, and $S_{\rm yr}$ to be the maximum value of the structure function evaluated for $\Delta t_{\rm 1yr}$ and $\Delta t_{\rm 2yr}$.  We also measure the intrinsic variability as defined by \citet{Sesar2007}, $\sigma_{int} = \sqrt{\Sigma^2 - \xi^2}$, where $\Sigma^2 = \frac{1}{n-1} \sum_{k=1}^{n}(m_{k}-\langle m \rangle)^2$ and $\xi^2 = \frac{1}{n} \sum_{k=1}^{n}\sigma(\langle m \rangle,k)^2$, and the maximum amplitude of variability, max($|\Delta m|$).

\subsection{UV Light Curve} \label{sec:flares}

\begin{figure}
\plotone{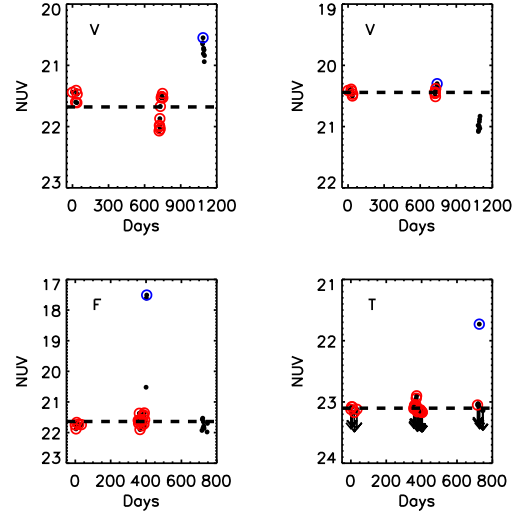}
\caption{Example light curves of sources flagged as stochastically variable ('V'), flares ('F'), and transient flares ('T').  Epochs $\ge 10$ days before the epoch of peak flux are circled in red, and the epoch of peak flux is circled in blue.  Thick dashed line indicates the mean flux $\ge 10$ days before the peak.
\label{fig:flare}
}
\end{figure}

In order to flag possible transient UV events that may be associated with a SN or TDE, we differentiate between stochastic variability and flaring variability.  We identify flaring UV variability as sources that show a constant flux $\ge$ 10 days before the peak of the light curve, and do not fade more than 2$\sigma$ below the faintest pre-peak magnitude (measured $\ge 10$ days before the peak).  This selection criteria is tailored to the $NUV$ rise times observed in SNe \citep{Gezari2008b, Brown2009, Gezari2010, Milne2010, Ofek2010} and TDE candidates \citep{Gezari2006, Gezari2008a, Gezari2009}.  We define constant pre-peak flux as a light curve with a reduced $\chi_{\nu}^{2}/ < 3~(\chi^{2}_{\nu} = \sum_{k=1}^{p} (m_{k} - \langle m \rangle)^{2}/(\sigma(\langle m \rangle ,k)^{2}/(n-1)$, where $n$ is the number of epochs $\ge 10$ days before the peak).  For those sources for which there are only upper limits $\ge $ 10 days before the peak, $\chi^{2}_{\nu}$ is set to 1.  Flaring sources with no detections before the peak are further labeled as transients.  Sources with $\chi_{\nu}^{2}/ \ge 3$, or that fade below 2$\sigma$ of the faintest magnitude measured $\ge 10$ days before the peak are labeled as stochastically variable.  We flag 116 flares, 145 transients, 595 stochastically variable sources, and remain with 222 sources with neither light-curve classification flag.  In Figure \ref{fig:flare} we show example light curves of sources flagged as stochastically variable ('V'), flares ('F'), and transients ('T').

\section{Host Properties} \label{sec:opt}

\subsection{Archival Optical Imaging Catalogs} \label{sec:archive}

We first characterize the host properties of the UV variable sources using archival optical $u$, $g$, $r$, $i$, and $z$ photometry and morphology from matches to the SDSS Photometric Catalog, Release 8 \citep{SDSSIII} ($m_{\rm lim} \sim 22$ mag), the CFHTLS Deep Fields D1, D2, and D3 ($m_{\rm lim} \sim 26.5$ mag) and Wide Fields W1, W3, and W4 ($m_{\rm lim} \sim 25$ mag) merged catalogs version T0005 \footnote{http://terapix.iap.fr/rubrique.php?id\_rubrique=252}, and the SWIRE ELAIS N1 and CDFS Region catalogs ($m_{\rm lim} \sim 24$ mag) \citep{Surace2004}.  For sources with matches in multiple catalogs, we use the match from the deepest catalog.  We convert the CFHTLS magnitudes to the SDSS system using the conversions in \cite{Regnault2009}, and the SWIRE Vega magnitudes to the SDSS system using the transformations measured for stellar objects available at the INT WFS web site \footnote{www.ast.cam.ac.uk/$\sim$wfcsur/technical/photom/colours}.  We then correct for Galactic extinction using the \citet{Schlegel1998} dust map values for $E(B-V)$ listed in Table \ref{tab1}.  Figure \ref{fig:fields} shows the overlap of the \textsl{GALEX} TDS fields with the available archival optical catalogs.  

\subsection{Pan-STARRS1 Medium Deep Survey} \label{sec:ps1}

The \textsl{GALEX} TDS fields overlap with the PS1 MDS fields MD01 (PS\_XMMLSS), MD02 (PS\_CDFS), MD04 (PS\_COSMOS), MD07 (PS\_GROTH), MD08 (PS\_ELAISN1), and MD09 (PS\_VVDS22H).  The \PS\ observations are obtained through a set of five broadband
filters, (\gps, \rps, \ips, \zps, and
\yps).  Further information on the passband shapes is described
in \cite{Stubbs2010}.  The PS1 MD fields are observed with a typical cadence in a given filter of 3 days, with an observation in the \gps\ and \rps\ bands on night one, in the \ips\ band on night two, and the \zps\ band on night three, with \yps-band observations during each of three nights on either side of the Full Moon.  Image differencing is performed on the nightly stacked images, reaching a typical 5$\sigma$ detection limit of $\sim 23.3$ mag per epoch in the \gps, \rps, \ips\ bands and $\sim 21.7$ mag in the \yps\ band.  Image difference detections from the PS1 Image Processing Pipeline (IPP; \cite{Magnier2006}) and an independent pipeline hosted by Harvard/CfA \citep{Rest2005} are internally distributed to the PS1 Science Consortium as transient alerts for visual inspection and classification.

\begin{figure}
\plottwo{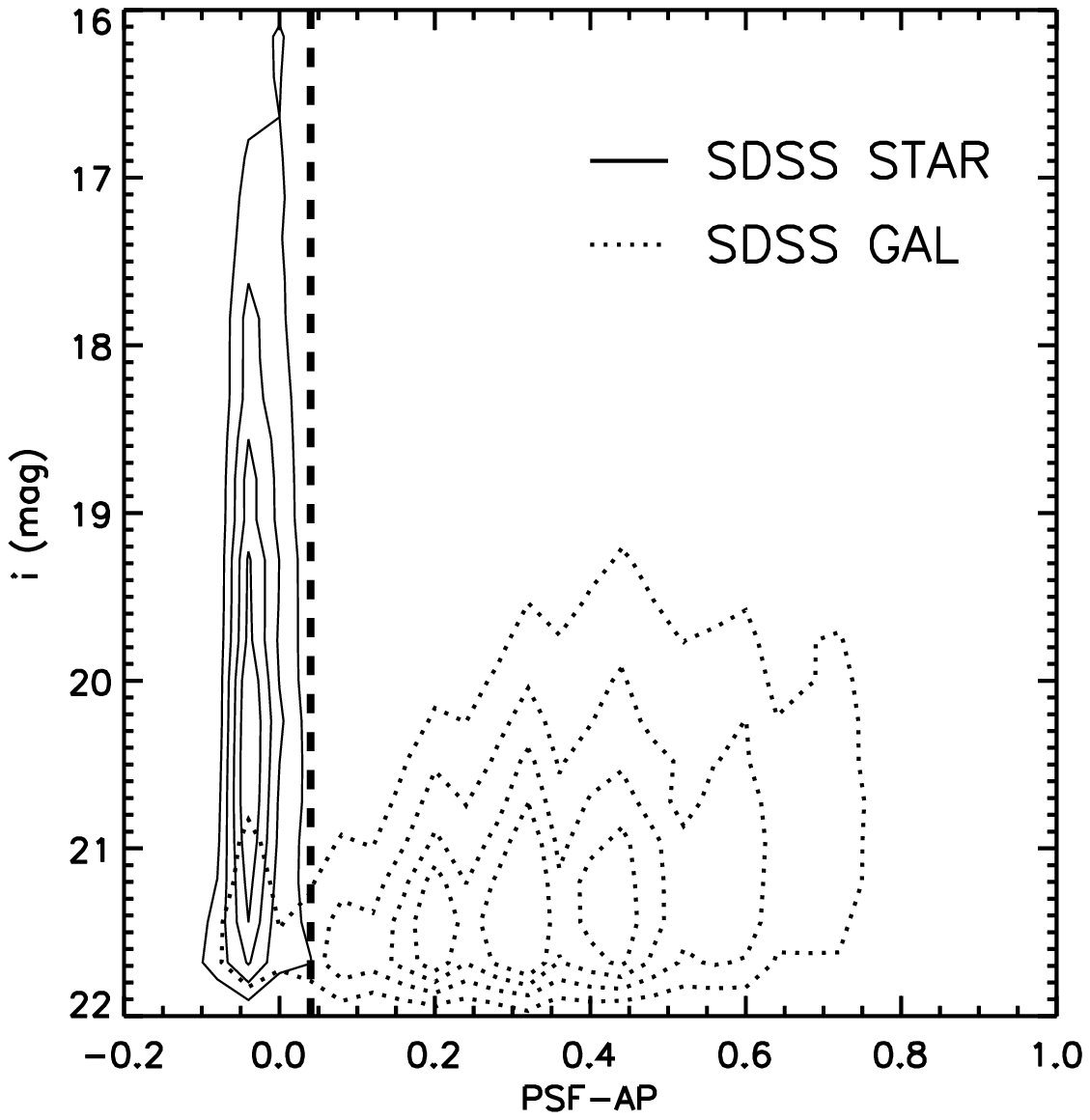}{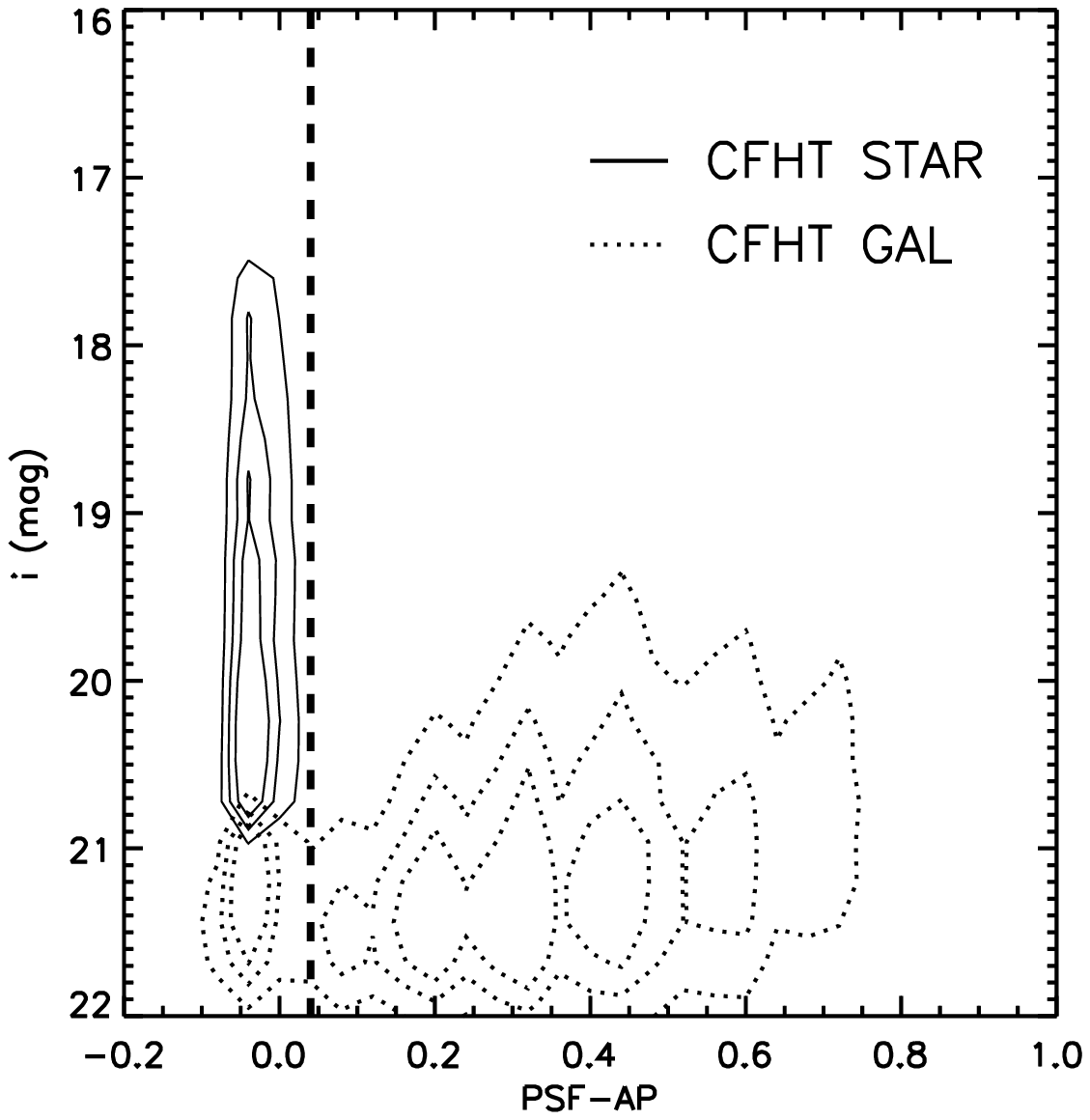}
\caption{Comparison of the star/galaxy separator used in the PS1 MD catalogs (thick dashed lines) to matches in the SDSS ({\it left}) and CFHT ({\it right}) archival catalogs in the PS\_GROTH field as a function of $i$-band magnitude.  
\label{fig:stargal}
}
\end{figure}

Deep stacks of the multi-epoch observations were generated to provide deep imaging with a 5$\sigma$ point-source limiting magnitude of $\sim$ 24.9, 24.7, 24.7, 24.3, 23.2 mag in the \gps, \rps, \ips, \zps, and \yps\ bands, respectively, and typical seeing (PSF FWHM) of $\sim 1.4, 1.3, 1.0, 1.0, 1.0$ arcsec in the 5 bands respectively.  
The magnitudes are in the ``natural'' \PS\  
system, $m=-2.5$log(flux)+$m'$, with a relative zeropoint adjustment $m'$ made in each band for each individual epoch \citep{Schlafly2012} before stacking to conform to the absolute flux calibration in the AB magnitude system \citep{Tonry2012}.  We convert the PS1 magnitudes to the SDSS system using the bandpass transformations measured for stellar SEDs in \citet{Tonry2012}, and correct for Galactic extinction using the \citet{Schlegel1998} dust map values for $E(B-V)$ listed in Table \ref{tab1}.
We obtain morphology information from the PS1 IPP output parameters in the \ips\ filter for the PSF magnitude ({\tt PSF\_INST\_MAG}), the aperture magnitude ({\tt PSF\_AP\_MAG}), and the PSF-weighted fraction of unmasked pixels {\tt PSF\_QF}, to define a point source or extended source as:
{\tt IF PSF\_INST\_MAG}$-${\tt PSF\_AP\_MAG} $< 0.04$ mag {\tt AND} {\tt PSF\_QF} $> 0.85$ {\tt THEN} class = pt
{\tt IF PSF\_INST\_MAG}$-${\tt PSF\_AP\_MAG} $> 0.04$ mag {\tt AND} {\tt PSF\_QF} $> 0.85$  {\tt THEN} class = ext
We calibrated these parameter cuts by comparing sources detected in both the PS1 MDS and archival optical catalogs.  Figure \ref{fig:stargal} shows the PS1 star/galaxy separation criteria for 110,804 sources detected in both PS1 and SDSS catalogs in the PS\_GROTH field, and for 169,461 sources detected in both PS1 and CFHT catalog in the PS\_GROTH field, with $i < 22$ mag, the faintest magnitude for 96\% of the optical hosts of the \textsl{GALEX} TDS sources, and the magnitude limit where all three
catalogs are complete.  Even though the CFHT catalogs are deeper than SDSS, they do not attempt to separate stars and galaxies for $i \apgt 21$ mag, and classify all sources fainter than this magnitude as point sources.  However, it is clear from both comparison plots, that the PS1 criterion of {\tt PSF\_INST\_MAG}$-${\tt PSF\_AP\_MAG} $< 0.04$ mag does an even better job of separating the locus of stars from galaxies than both catalogs down to $i \sim 22$ mag.

\subsection{Archival Redshift and X-ray Catalogs}

We also take advantage of the many archival X-ray and spectroscopic catalogs available from the overlap of the \textsl{GALEX} TDS survey with legacy survey fields.  In the PS\_CDFS field we us X-ray catalogs from the 0.3 deg$^{2}$ Chandra Extended CDFS survey \citep{Giacconi2002, Lehmer2005, Virani2006}, and redshift catalogs from the VIMOS VLT Deep Survey (VVDS) \citep{LeFevre2004}, and a compilation of redshift catalogs from GOODS and SWIRE \footnote{http://www.eso.org/$\sim$arettura/CDFS\_master/index.html}.  In the PS\_XMMLSS field we use X-ray catalogs from the 5.5 deg$^{2}$ XMM-LSS survey \citep{Chiappetti2005, Pierre2007}, and redshift catalogs from VVDS \citep{LeFevre2005}.  In the PS\_COSMOS field, we use X-ray catalogs from the 1.9 deg$^{2}$ XMM-Newton Wide-Field Survey \citep{Hasinger2007} and the 0.9 deg$^{2}$ Chandra COSMOS survey \citep{Elvis2009}, and redshifts from the Magellan COSMOS AGN survey \citep{Trump2007, Trump2009}, the VLT zCOSMOS bright catalog \citep{Lilly2007, Lilly2009}, and the Chandra COSMOS Survey catalog \citep{Civano2012}.  In the PS\_GROTH field we use X-ray catalogs from the 0.67 deg$^{2}$Chandra Extended Groth Strip \citep{Nandra2005, Laird2009} and redshift catalogs from the DEEP2 Galaxy Redshift Survey \citep{Newman2012}.  For PS\_ELAISN1 we use the X-ray catalog from the 0.08 deg$^{2}$ Chandra ELAIS-N1 deep X-ray survey \citep{Manners2003}.   
Figure \ref{fig:fields} shows the overlap of the \textsl{GALEX} TDS fields with the archival X-ray surveys.  Finally, we also match the sources with the ROSAT All-Sky Bright Source and All-Sky Survey Faint Source catalogs \citep{Voges1999, Voges2000}.

\section{Classification} \label{sec:class}

We classify the \textsl{GALEX} TDS sources using a combination of optical host photometry and morphology, UV variability statistics, and matches with archival X-ray and redshift catalogs.  Table \ref{tab:class} summarizes the sequence of steps we use to classify the sources, which we describe in detail below.

\subsection{Cross-Match with Optical Catalogs}
We first cross-matched our 1078 \textsl{GALEX} TDS sources with the archival $u,g,r,i,z$ catalogs described in \S\ref{sec:archive} with a matching radius of 3 arcsec.  This radius is recommended for matches between \textsl{GALEX} and ground-based optical catalogs \citep{Budavari2008}, and corresponds to a spurious match rate of only $1-2\%$ at the high Galactic latitudes of the \textsl{GALEX} TDS fields \citep{Bianchi2011}. However, we found that there was a population of ``orphans'' (no optical match within 3 arcsec) that were detected in their $NUV$ low-state, and had a match between $3-4$ arcsec with an optically identified quasar.  Given the strong likelihood that these are real matches, we increased our matching radius to 4 arcsec.  We use the star/galaxy classifications from the catalogs to label sources as point sources (pt) or extended sources (ext).  This results in 878/1078 optical matches (81\%), with the majority of sources without matches in PS\_CDFS, which is only partially covered by the archival catalogs. 

\begin{figure}
\plottwo{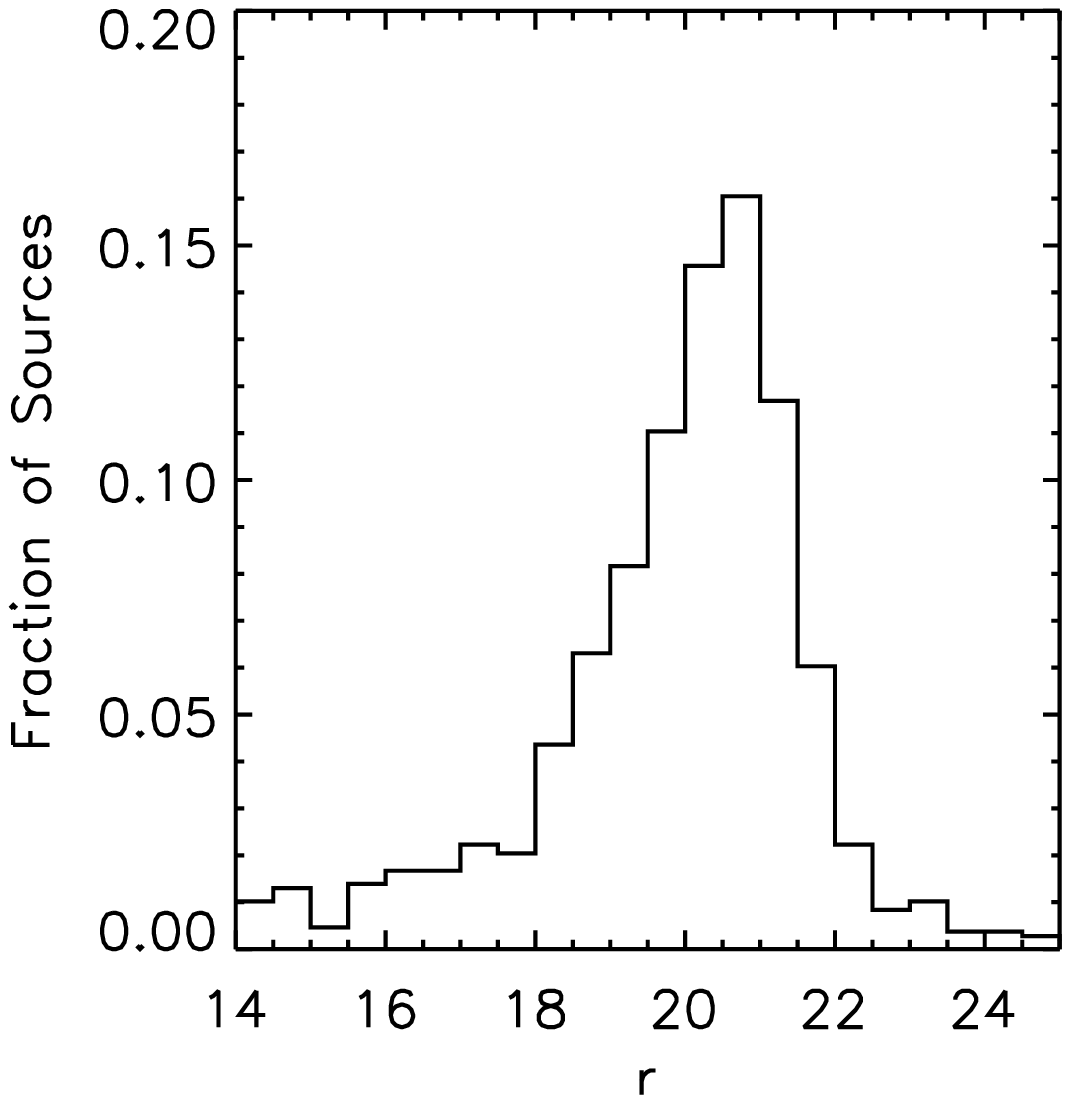}{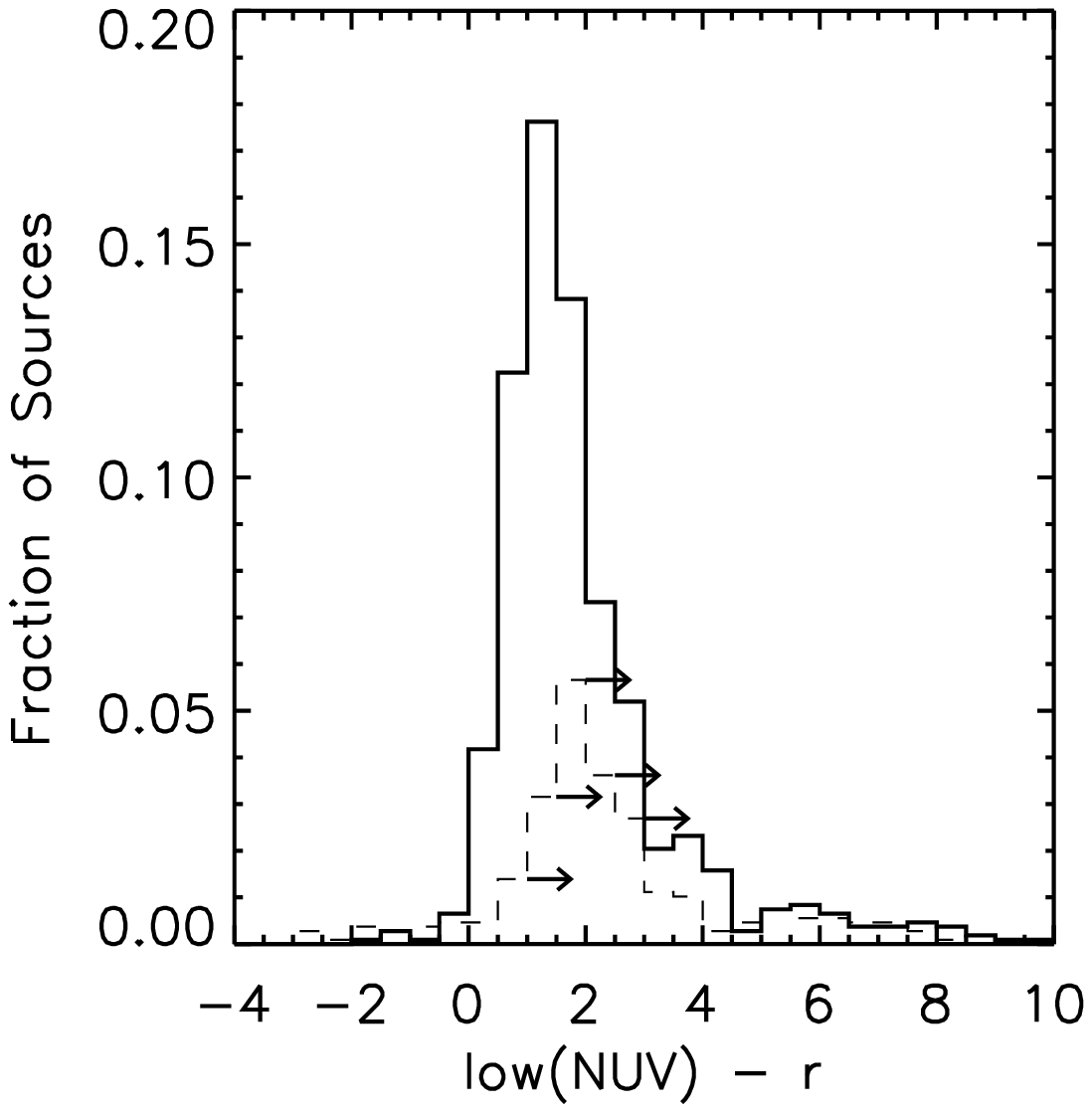}
\caption{{\it Left}:  Histogram of $r$ magnitudes of optical matches.  {\it Right}: Histogram of $NUV-r$ colors of optical matches for those sources detected in the $NUV$ during their low-state (solid lines), and those sources with upper limits during their low-state (dashed lines).  
\label{fig:opthist}
}
\end{figure}

We then match the sources that do not have archival optical matches to the PS1 MDS catalog described in \ref{sec:ps1}, this increases the number of sources with optical photometry and morphology (albeit without the $u$ band) to 1057/1078 (98\%).  Figure \ref{fig:opthist} shows a histogram of the $r$-band magnitude of the optical hosts, and the $NUV-r$ colors of the \textsl{GALEX} TDS sources in their low-state.  The optical hosts have a distribution that peaks at $r \sim 21$ mag, over 3 magnitudes brighter than the detection limit of PS1 MDS, and $NUV - r \sim 1$ mag. Sources not detected in their $NUV$ low-state are shown as upper limits in the $NUV-r$ color histogram, and peak at $NUV - r > 2$ mag.

\subsection{Orphans} \label{sec:orphans}
We visually inspected the PS1 stack images at the locations of the 21 sources with no optical matches, and confirm that they are true orphan events.  Furthermore, all of the orphans are undetected in their low-state in the $NUV$, with upper limits of $NUV > (22.3-23.1$) mag.  Thus the orphan hosts are likely distant stars or faint galaxies (i.e., dwarf galaxies or high-redshift galaxies) that are undetected during their low-state in the optical {\it and} $NUV$.    

\subsection{Color and Morphology Cuts} \label{sec:color}

\begin{figure}
\plotone{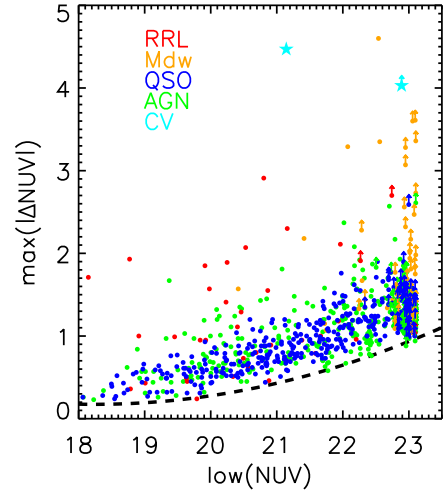}
\caption{Maximum $NUV$ variability amplitude as a function of low-state $NUV$ magnitude for sources classified as RR Lyrae (red), M dwarfs (yellow), quasars (blue) AGN (green), and CVs (cyan).  Dashed line shows the median 5$\sigma$ error selection function used to select the variable sources.
\label{fig:delmax}
}
\end{figure}

We first use the color and morphology of the optical hosts to classify the \textsl{GALEX} TDS 5$\sigma$ UV variable sources.  We define quasars as sources with optical point-source hosts with 

\begin{eqnarray}
u-g < 0.7 \\
-0.1 < g-r < 1.0 \nonumber 
\end{eqnarray}

\noindent  in order to avoid the stellar locus and white dwarfs \citep{Richards2002}.   Note that this color selection can be contaminated by catacylismic variable stars (CVs), which overlap in color-color space with the quasar sample.  Indeed, two of the sources classified by color as quasars are in fact spectroscopically confirmed CVs (VVDS22H\_MOS05-05 and ELAISN1\_MOS15-02).  VVDS22H\_MOS05-05 is ROTSE3 J221519.8-003257.2, a confirmed cataclysmic variables with a dwarf-nova type spectrum.  We observed ELAISN1\_MOS15-02 with the APO 3.5m telescope Dual Imaging Spectrograph (DIS) on 2011 May 3 and detected broad Balmer emission lines from a Galactic source, characteristic of a CV/dwarf nova spectrum.  While these sources stood out easily because of their extreme magnitude of variability of $|\Delta m| > 4$ mag, shown in Figure \ref{fig:delmax}, there may be lower amplitude CV events that are still hiding in our quasar sample.  However, given the low surface density of CVs relative to quasars in the sky \citep{Szkody2011}, the expected contamination rate is consistent with the two CVs identified. 

We define RR Lyrae stars as sources with optical point source hosts with 

\begin{eqnarray}
0.75 < u-g < 1.45 \\
-0.25 < g-r < 0.4 \nonumber \\
-0.2 < r-i < 0.2 \nonumber \\
-0.3 < i-z < 0.3 \nonumber 
\end{eqnarray}

\noindent \citep{Sesar2010}.   Note that the color cuts for quasars and RR Lyrae require the $u$-band, which is not available for sources with PS1-only matches.  However, we define M dwarf stars as point sources with 

\begin{eqnarray}
r-i > 0.42 \\
i-z > 0.24 \nonumber \\
g < 22.2 \nonumber \\
r < 22.2 \nonumber \\
i < 21.3
\end{eqnarray}

\begin{figure}
\plotone{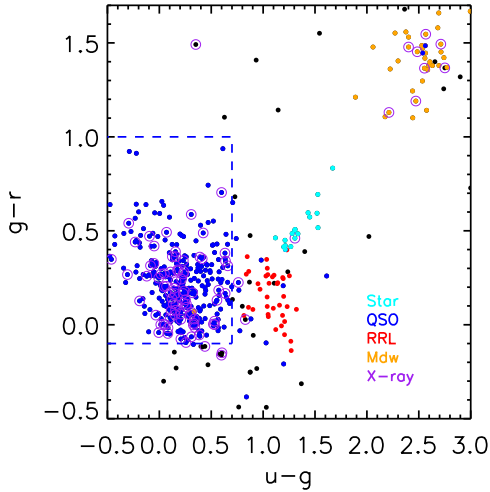}
\caption{Colors of archival optical matches to UV variable sources with point-like optical hosts (black points).  Dashed blue line shows the region in color-color space used to define quasars from optical colors and morphology alone.  Sources with classification are color coded as quasars in blue, RR Lyrae in red, and M dwarf stars in yellow, and main stellar locus stars in cyan.  Sources with archival X-ray matches are circled in purple.
\label{fig:colors_pt}
}
\end{figure}

\noindent \citep{West2011}, which does not require $u$-band data.  We classify stars on the main stellar locus as those with 

\begin{eqnarray}
1.0 < u-g < 2.25 \\
0.4 < g-r < 1.0 \nonumber \\
-0.2 < r-i < 2.0 \nonumber \\
-0.3 < i-z < 1.0 \nonumber 
\end{eqnarray}

\noindent modified from \citet{Yanny2009}. This color and morphology selection results in \rr RR Lyrae, \mm M-dwarf flare stars, \ms stars, and 325 quasars.  Figure \ref{fig:colors_pt} shows the optical color-color diagram of the sources with optical point-source hosts and $u$-band data, and their classifications as quasars, RR Lyrae, M dwarfs, and stars. 

\subsection{UV Variability Cuts} \label{sec:varcut}

\begin{figure}
\plotone{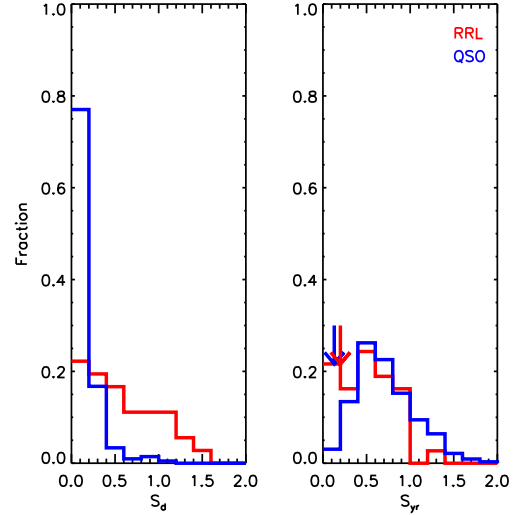}
\caption{Histogram of the $NUV$ structure function of sources classified as RR Lyrae (red) and quasars (blue) on a timescale of days and years.  Red arrow and Blue arrow indicate the mean of S$_{\rm yr}$ measured in the SDSS $r$-band for RR Lyrae and quasars, respectively from \cite{Schmidt2010}. 
\label{fig:struct}
}
\end{figure}

\begin{figure}
\plotone{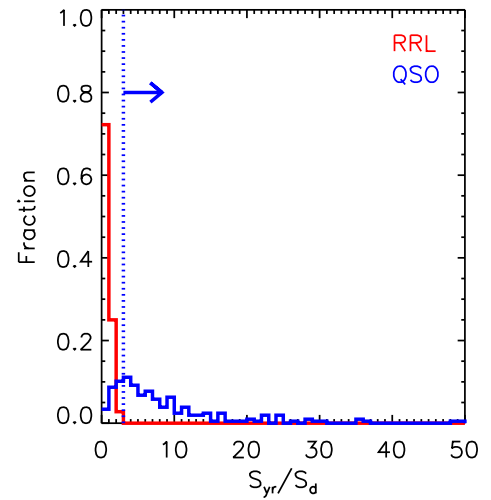}
\caption{Histogram of the ratio of the $NUV$ structure function on timescales of years and days for sources classified as RR Lyrae (red) and quasars (blue).  Dotted line shows the selection criteria of $S_{\rm y}/S_{\rm d} \ge 3$ used to classify sources with optical point source hosts as quasars.
\label{fig:struct_ratio}
}
\end{figure}

\begin{figure}
\plotone{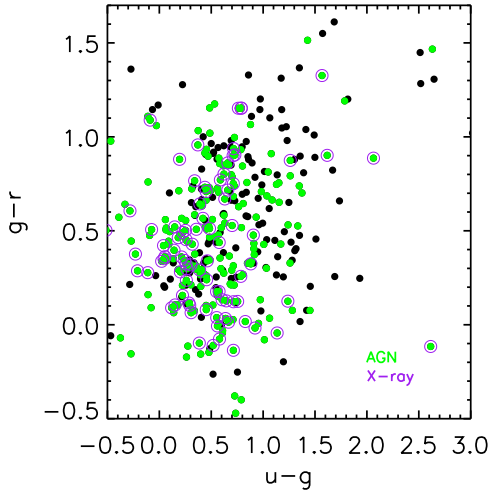}
\caption{Colors of archival optical matches to UV variables with extended optical hosts.  Sources classified as AGN (by either stochastic UV variability, archival spectra, and/or an X-ray match) are color coded in green.  Sources with matches with archival X-ray matches are circled in purple.  Note that all galaxy sources with an archival X-ray match are defined as AGNs.
\label{fig:colors_ext}
}
\end{figure}

Figure \ref{fig:struct} shows the structure function on timescales of days and years described in \S\ref{sec:stats} for the sources classified as RR Lyrae and quasars in \S\ref{sec:color}.  
 In Figure \ref{fig:struct_ratio} we show the ratio of the $NUV$ structure function on timescales of years to days ($S_{\rm yr}/S_{\rm d}$).  While quasars demonstrate a wide range of $S_{\rm yr}/S_{\rm d}$, all RR Lyrae have $S_{\rm yr}/S_{\rm d} < 3$.  We use this UV variability property to relax our color constraints, and increase our photometric sample of quasars to all sources with optical point-source hosts with $S_{\rm yr}/S_{\rm d} \ge 3$.  This is equivalent to a structure function power-law exponent cut of $\gamma > 0.2 $, where $S(\Delta t) \propto \Delta t^{\gamma}$ \citep{Hook1994, VandenBerk2004, Schmidt2010}.  This structure-function ratio selection results in the classification of another 30 quasars.  Two additional sources with optical point-source hosts have archival quasar spectra, resulting in a final quasar sample of \qso.   
We define active galactic nuclei (AGN), as sources with optically extended hosts that show stochastic UV variability (see \S\ref{sec:flares}), have an X-ray catalog match, and/or an archival spectroscopic classification.  Figure \ref{fig:colors_ext} shows the optical color-color diagram of the sources with optically extended hosts, and those classified as AGN.  This results in a sample of \agn AGN.  We also add archival spectroscopic classifications for 6 stars.  This yields a total of 776/1078 (72\%) sources classified as an active galaxy (quasar or AGN) or variable star.

\subsection{X-ray Sources} \label{sec:xray}

The archival X-ray catalogs overlap with $\sim 8.45$ deg$^{2}$ of the \textsl{GALEX} TDS survey area.  Within this area, 81/89 quasars, 92/105 AGN, and 8/9 M-dwarf stars are detected in the X-rays.  In addition, there are 9 optical point sources with X-ray matches that are likely quasars and M dwarfs just outside the quasar and M-dwarf color-color selection regions.  UV variability selection appears to be selecting a similar population of active galaxies and M dwarfs as X-ray detection, since $\sim$ 90\% of the UV variability-selected active galaxy and M dwarf sample is also detected in the X-rays.  However, only 2\% of all the X-ray sources (the majority of which are active galaxies) are detected as UV variable at the selection threshold of the \textsl{GALEX} TDS catalog.

\subsection{GUVV Catalog} \label{sec:guvv}

We also cross-match our \textsl{GALEX} TDS sample with the first and second \textsl{GALEX} Ultraviolet Variability Catalogs (GUVV-1 \& GUVV-2) from \cite{Welsh2005} and \citet{Wheatley2008}.  These catalogs include 894 UV variable sources ($\Delta m > 0.6$ mag in the $NUV$) selected from an analysis of archival \textsl{GALEX} AIS, MIS, DIS, and Guest Investigator (GI) fields with repeated observations.  With a cross-matching radius of 4 arcsec, we find a match with 36 GUVV sources.  For the 15 matches that have GUVV classifications, are all classified by GUVV as active galaxies (AGN or quasars), which are in agreement with our \textsl{GALEX} TDS classifications.  Of the 21 matches without GUVV classifications, we find 5 sources classified by \textsl{GALEX} TDS as RR Lyrae, 9 classified as active galaxies (AGN or quasars), 6 with optical point-source hosts, and 1 with a galaxy host. 

\subsection{Unclassified Sources}

\begin{figure}
\plotone{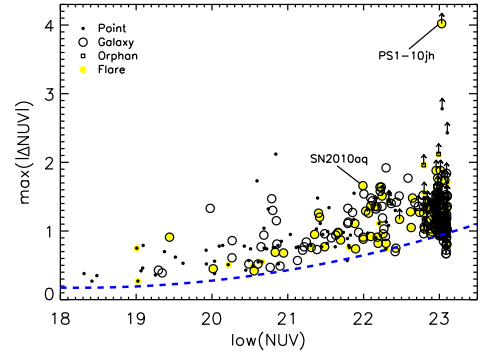}
\caption{Maximum $NUV$ variability amplitude for sources without a classification.  Dashed blue line shows the median 5$\sigma$ variability selection function as a function of mean magnitude.  The nature of the optical host is indicated (point source, galaxy, or orphan), and flaring sources (including transients) are marked in yellow.
\label{fig:delmax_unclass}
}
\end{figure}

The remaining 302 unclassified sources include 91 with optical point-source hosts, which are likely stars, quasars with non-standard colors (high-redshift or reddened), or unresolved galaxies, 190 with galaxy hosts, and 21 orphans. The 190 galaxy hosts may either be faint AGN with poorly constrained UV light curves, or hosts of UV-bright extragalactic transients.  In Figure \ref{fig:delmax_unclass} we show the maximum $|\Delta m|$ in the $NUV$ as a function of low-state $NUV$ magnitude of the remaining unclassified sources.
The unclassified UV source with the most extreme amplitude, ELAISN1\_MOS15-09 with $|\Delta m| > 4.2$ mag, was spectroscopically confirmed to be from the nucleus of an inactive galaxy at $z = 0.1696$, and its UV/optical flare detected by \textsl{GALEX} TDS and PS1 MDS was attributed to the tidal disruption of a star around a supermassive black hole \citep{Gezari2012}.  Also in this sample is a UV transient spectroscopically confirmed to be a Type IIP SN 2010aq at z=0.086 (COSMOS\_MOS26-29), whose UV/optical light curve from \textsl{GALEX} TDS and PS1 MDS was fitted with early emission following SN shock breakout in a red supergiant star \citep{Gezari2010}.  Both of these spectroscopically classified extragalactic transients are labled in Figure \ref{fig:delmax_unclass}.

Our 5$\sigma$ selection criteria translates to a limiting sensitivity to transients in a host galaxy with a magnitude $m_{\rm host}$ of a magnitude of
\begin{equation}
m_{\rm trans} = -2.5 \log (10^{\frac{m_{\rm host} - 5\sigma(m_{\rm host})}{-2.5}} - 10^{\frac{m_{\rm host}}{-2.5}}),
\end{equation}
which ranges from $m_{\rm trans} \sim 20.0$ mag for $m_{\rm host} = 18$ mag to $m_{\rm trans} \sim 22.7$ mag for $m_{\rm host}=23$ mag.  Thus, our variability selection threshold is less sensitive to transients in host galaxies with bright $NUV$ fluxes.  On the red sequence of galaxies, where $M_{\rm NUV} \approx -14.5$ \citep{Wyder2007}, this selection effect is not as much of an issue, since already for $z > 0.05$ one gets $m_{\rm host}$ $>$ 22 mag.  However, star-forming galaxies on the blue sequence are 2.5 mag brighter in the $NUV$, and thus the host galaxy brightness can be a factor in reducing the senstivity to faint transients.  For example, our \textsl{GALEX} TDS 5$\sigma$ sample does not include SN 2009kf, a luminous Type IIP SN in a star-forming galaxy at $z=0.182$ which we reported our \textsl{GALEX} TDS detection of in \cite{Botticella2010}.  This source varied at only the 4.25$\sigma$ level in the $NUV$ during its peak.  However, because this transient was selected from a spatial and temporal coincidence with a PS1 transient alert, we could lower our threshold for variability selection in the UV.  The systematic selection of SN and TDE candidates from the joint \textsl{GALEX} TDS and PS1 MDS transient detections will be presented in future papers.
 

\section{Discussion} \label{sec:disc}

\subsection{Classification Demographics}
\begin{figure}
\plotone{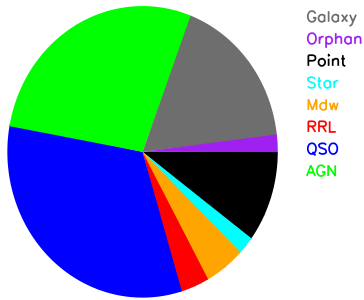}
\caption{Pie chart of \textsl{GALEX} TDS classifications: sources classified as stars (Star), M dwarfs (Mdw), RR Lyrae (RRL), quasars (QSO), and active galactic nuclei (AGN), and sources with no classification with galaxy hosts (Galaxy), no host (Orphan), and point-source hosts (Point).
\label{fig:pie}
}
\end{figure}

Figure \ref{fig:pie} shows a pie diagram of the source classifications.  Out of the total of 1078 \textsl{GALEX} TDS sources, 62\% are classified as actively accreting supermassive black holes (quasars or AGN), and 10\% as variable and flaring stars (including RR Lyrae, M dwarfs, and CVs).  Note that the relative fraction of the different classes of sources is sensitive to both their intrinsic magnitude distribution, and the magnitude-dependent variability selection function of the sample.  Table \ref{tab2} gives the \textsl{GALEX} TDS catalog, sorted by decreasing $NUV$ amplitude, with the \textsl{GALEX} ID, R.A., Dec, low-state $NUV$ magnitude, maximum amplitude of $NUV$ variability ($|\Delta m_{\rm max}|$), intrinsic variability ($\sigma_{int}$), the structure function on day ($S_{\rm d}$) and year ($S_{\rm y}$) timescales, the characteristics of the $NUV$ light curve: flaring (F) or stochastically variable (V), the morphology of the matching optical host: point-source (pt) or extended (ext), the color classification of the matching optical host: RR Lyrae (RRL), M dwarf star (Mdw), star (Star), or quasar (QSO), the archival redshift, an X mark if there is a match with an archival X-ray source, and finally the \textsl{GALEX} TDS classification: RR Lyrae (RRL), M dwarf star (Mdw), star (star), quasar (QSO), AGN, UV flaring source or UV transient source with galaxy host (Galaxy Flare or Galaxy Trans), UV flaring source or transient source with point-source optical host (Point Flare or Point Trans) or UV flaring source or transient source with orphan optical host (Orphan Flare or Orphan Trans), stochastically variable source with a point-source optical host (Point Var), stochastically variable orphan (Orphan Var), or none of the above (?).  

\begin{figure}
\plottwo{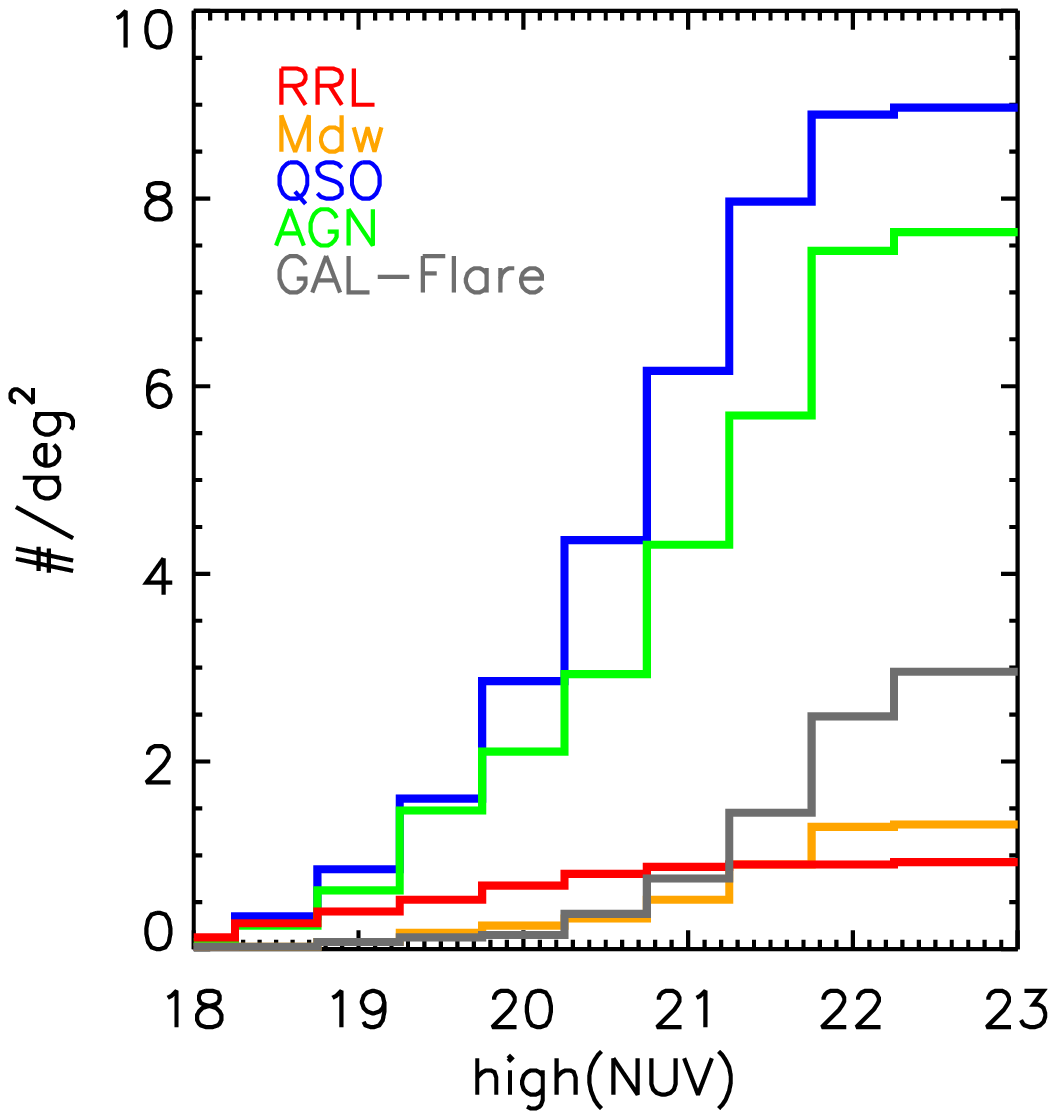}{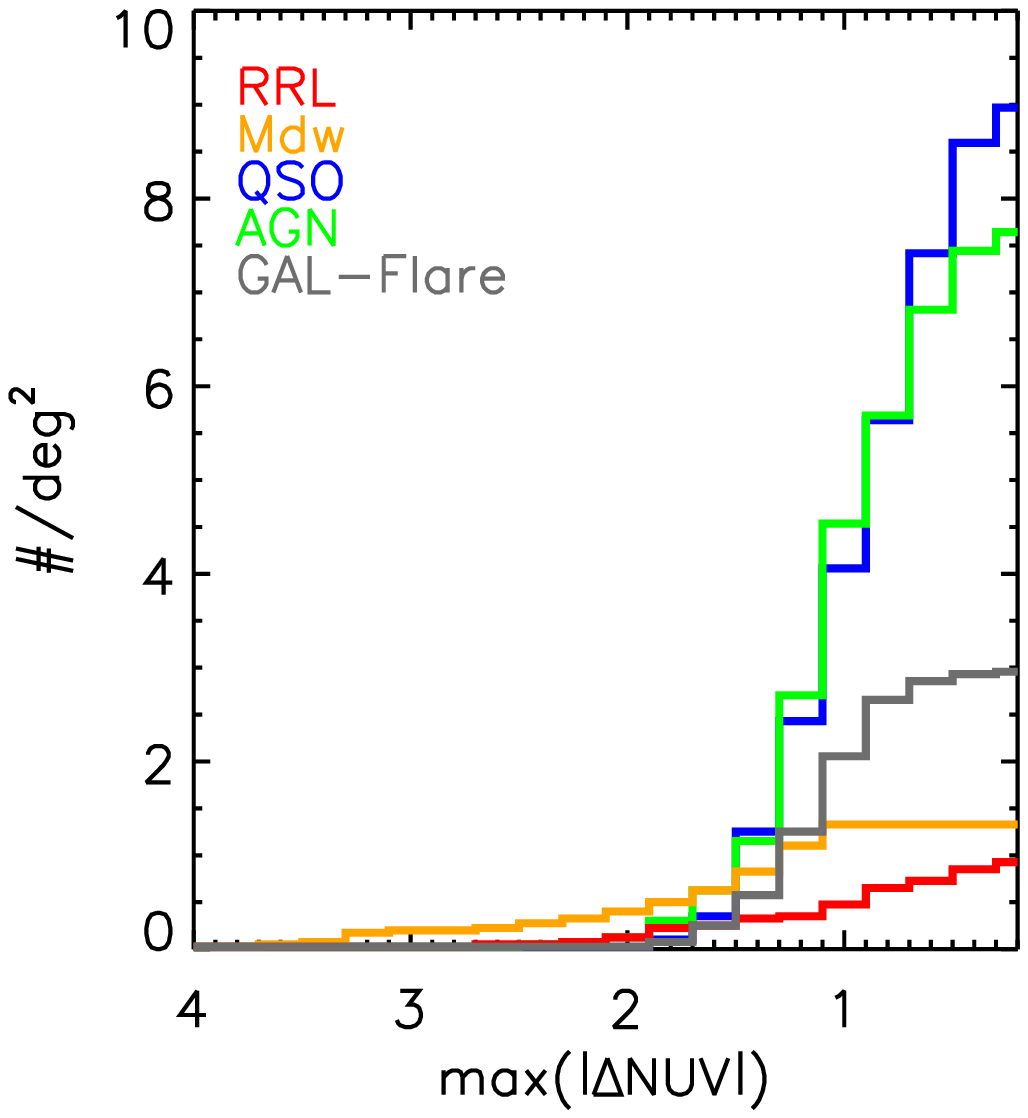}
\caption{Cummulative distribution of surface density of UV variable sources with \textsl{GALEX} TDS classifications: M dwarfs (Mdw), RR Lyrae (RRL), quasars (QSO), active galactic nuclei (AGN), and extragalactic transients (GAL-Flare) as a function of peak magnitude (high(NUV)) ({\it left}) and maximum amplitude (max($|\Delta NUV|$)) ({\it right}).
\label{fig:dens_mag}
}
\end{figure}

In Figure \ref{fig:dens_mag} we show the cummulative surface density distribution of classified UV variable sources as a function of peak magnitude (high(NUV)) and maximum amplitude (max($|\Delta NUV|$)).  For the variable UV sources, these correspond to total surface densities of $8.0 \pm 3.1$, $7.7 \pm 5.8$, and $1.8 \pm 1.0$ deg$^{-2}$ for quasars, AGNs, and RR Lyrae, respectively.  For the transient source, we can calculate a total surface density rate, $\#/(area \times t_{\rm eff})$, where $t_{\rm eff}$ is the effective survey time at the cadence that matches the characteristic timescale of the transient. For extragalactic transients such as young SNe and TDEs, which vary on a timescale of days, we use the time intervals for which the fields were observed with a cadence of $2.0 \pm 0.5$ days.  If we include all flaring or transient \textsl{GALEX} TDS sources with a galaxy host, this yields a surface density rate of $52 \pm 38$ deg$^{-2}$ yr$^{-1}$ for extragalactic transients.  For M dwarfs which vary on timescales shorter than an individual observation, we use the total exposure time for each epoch.  If we assume a survey with a cadence of 2 days and $t_{\rm exp} = 1.5$ ksec, this translates to a surface density rate for M dwarfs of $15 \pm 10$ deg$^{-2}$ yr$^{-1}$.

\subsection{UV Variability Properties of Classified Sources}

\begin{figure}
\plotone{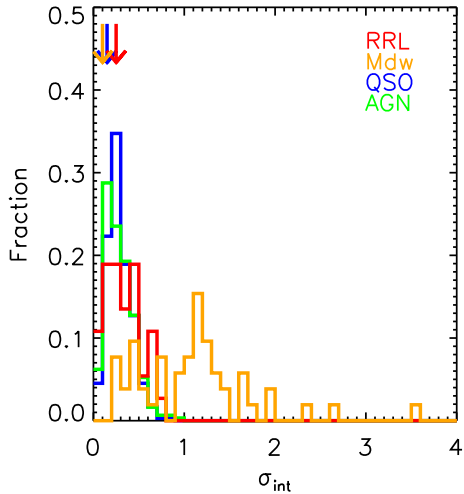}
\caption{Intrinsic $NUV$ variability as a function of low-state $NUV$ magnitude for sources classified as RR Lyrae (red), M dwarfs (yellow), quasars (blue) and AGN (green).  Arrows shows the median $\sigma_{\rm int}$ measured in the optical for quasars (blue arrow), RR Lyrae (red arrow), and M dwarfs (orange arrow) from \cite{Sesar2007}.
\label{fig:sig_int}
}
\end{figure}

Various optical studies of rest-frame UV variability in high redshift quasars have demonstrated that the variability of AGN increases with decreasing rest wavelength \citep{diClemente1996, VandenBerk2004, Wilhite2005}.  In Figure \ref{fig:sig_int}, we show histograms of $\sigma_{int}$ for the UV variable sources with classifications.  Quasars show a distribution of $\sigma_{\rm int}$ with a mean that is $\sim 2$ times larger than measured at optical wavelengths from the SDSS Stripe 82 sample from \cite{Sesar2007}.  This effect is even more pronounced in the magnitude of the structure function on years timescales (S$_{\rm y}$), which has a mean that is $5$ times larger than the mean measured in the $r$-band ($\lambda_{eff}=6231)$ from \cite{Schmidt2010}.  This trend is consistent with the wavelength dependent rise in variability amplitude observed in the structure function for quasars in the rest-frame UV \citep{VandenBerk2004} and observed UV \citep{Welsh2011}.  

The fact that AGN become bluer during high states of flux \citep{Giveon1999, Geha2003, Gezari2008a} has been attributed to increases in the characteristic temperature of the accretion disk in response to increases in the mass accretion rate \citep{Pereyra2006, LiCao2008}.  However, \citet{Schmidt2012} argue that the color variability observed in {\it individual} quasars in their SDSS Stripe 82 sample is stronger than expected from just varying the accretion rate ($\dot M$) in accretion disk models.  In a future study, we will use simultaneous UV and optical light curves from \textsl{GALEX} TDS and PS1 MDS for our \qso individual quasars to test this result with a larger dynamic range in wavelength.

\begin{figure}
\plotone{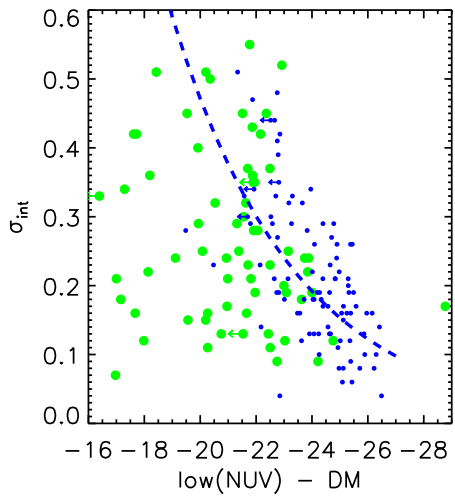}
\caption{Intrinsic $NUV$ variability as a function of low-state $NUV$ absolute magnitude, $M_{NUV}$ = low($NUV) - DM$ where $DM$ is the distance modulus, for the subsample of quasars (blue dots) and AGN (green circles)  with catalog redshifts.  Dashed blue line shows the fit to the quasars to $\log (\sigma_{int}) \propto \frac{\beta}{2.5}M$.  The spectroscopically classified extragalactic transients (TDE PS1-10jh and SN 2010aq) are labeled.
\label{fig:sig_abs}
}
\end{figure}

For the subsample of 95 quasars with archival redshifts ($z_{\rm mean} = 1.26$, $\sigma_{z} = 0.39$), in Figure \ref{fig:sig_abs} we plot $\sigma_{int}$ vs. the low-state $NUV$ absolute magnitude, and find a steep negative correlation fitted by $\log(\sigma_{\rm int}) = (1.6 \pm 0.1) + \frac{\beta}{2.5} M_{NUV}$, where $\beta = 0.24 \pm 0.04$, in excellent agreement with the trend for increased variability in lower luminosity quasars seen from optical observations with $\beta = 0.246 \pm 0.005$ \citep{VandenBerk2004}, and shallower than expected for a Poissonian process which has $\beta = 0.5$ \citep{CidFernandes2000}.  We also show the subset of 68 AGN with archival redshifts ($z_{\rm mean} = 0.64$, $\sigma_{z} = 0.55$), which clearly do not show a relation between $\sigma_{\rm int}$ and low-state $NUV$ absolute magnitude.  This is most likely a result of dilution of the variability amplitude from the contribution of the host galaxy in the $NUV$.

The largest values of $|\Delta m|$ (plotted in Figure \ref{fig:delmax}) are found in RR Lyrae and M dwarfs, with a tail of large amplitude variations reaching up to $|\Delta m| = 2.9$ mag in RR Lyrae and up to $|\Delta m| = 4.6$ mag in M dwarfs.  In the optical, the RR Lyrae structure function is very weakly dependent on timescale, with an amplitude of $0.1-0.2$ mag \citet{Schmidt2010}.  The $NUV$ structure function also shows a weak dependence of amplitude on timescale when comparing the structure function on days to years timescales, however, with an amplitude that is $\sim$3 times larger in the $NUV$ than in the optical.  This wavelength dependence on variability amplitude can be explained if the variability is driven by variations in surface temperature from pulsations of the stellar envelope, where higher states of flux are associated with higher temperatures \citep{Sesar2007}.

\section{Conclusions} \label{sec:conc}

We provide a catalog of over a thousand UV variable sources and their classifications based on optical host properties, UV variability behavior, and cross-matches with archival X-ray and redshift catalogs.  This yields a sample of \mm M dwarfs, \rr RR Lyraes, \qso quasars, and \agn AGN.  We find median intrinsic UV variability amplitudes in RR Lyrae and quasars that are factors of $> 3$ larger than at optical wavelengths, consistent with the expectation for higher temperatures during higher states of flux.  The regular cadence and wide area of \textsl{GALEX} TDS enables us to systematically discover persistent and transient (i.e. tidal disruption of a star) accreting supermassive black holes over wide fields of view, study the contemporaneous UV and optical variability of variable stars, and catch young core-collapse supernovae within the first days after explosion.  The overlap of the \textsl{GALEX} TDS with PS1 MDS and multiwavelength legacy survey fields will continue to be helpful for classifying transient sources in these heavily observed fields.  We also measure the surface densities of variable sources and the surface density rates for transients as a function of class in the UV for the first time.

With \textsl{GALEX} TDS we are only scratching the surface of UV variability. 
Our 5$\sigma$ sample of 1078 sources is less than 0.3\% of the 419,152 UV sources in the field (with an average density of $1.1\times10^{4}$ UV sources per square degree down to $m_{\rm lim} = 23$ mag).  Looking to the future, the discovery rate for UV variable sources and UV transients could increase by {\it several} orders of magnitude with the launch of a space-based UV mission with a wide field of view (several deg$^{2}$), a survey strategy of daily cadence observations over $\sim 100$ deg$^{2}$, and detectors with an order of magnitude improved photometric precision relative to \textsl{GALEX}.  In the optical sky, 90\% of quasars vary with $\sigma_{\rm int} > 0.03$ mag on the timescales of years \citep{Sesar2007}.  Given the factor of $\sim 2$ larger $\sigma_{\rm int}$ observed for quasars in the $NUV$, one could achieve a nearly complete sample of low-redshift quasars with photometric errors of $\sigma(m) \sim 0.01$ mag.  In coordination with a ground-based optical survey, such as Pan-STARRS2 \citep{Burgett2012} or LSST \footnote{lsst.org/lsst/science/overview}, this could yield the simultaneous UV and optical detection of $\approx 10^{4}$ variable quasars and $\approx 10^{3}$ RR Lyrae and M dwarfs, as well as increase the discovery rate of UV-bright extragalactic transients (young SNe and TDEs) by a factor of $\approx 100$. 

\acknowledgements

We thank the anonymous referee for their constructive comments which improved the paper.  \textsl{GALEX} (Galaxy Evolution Explorer) is a NASA Small Explorer, launched in 2003 April. We gratefully acknowledge NASA’s support for construction, operation, and science analysis for the \textsl{GALEX} mission, developed in cooperation with the Centre National d’Etudes Spatiales of France and the Korean Ministry of Science and Technology.  The Pan-STARRS1 survey has been made possible through contributions of the Institute for Astronomy, the University of Hawaii, the Pan-STARRS Project Office, the Max-Planck Society and its participating institutes, the Max Planck Institute for Astronomy, Heidelberg and the Max Planck Institute for Extraterrestrial Physics, Garching, The Johns Hopkins University, Durham University, the University of Edinburgh, Queen's University Belfast, the Harvard-Smithsonian Center for Astrophysics, and the Las Cumbres Observatory Global Telescope Network, Incorporated, the National Central University of Taiwan, and the National Aeronautics and Space Administration under grant No. NNX08AR22G issued through the Planetary Science Division of the NASA Science Mission Directorate.





\clearpage

\begin{deluxetable}{lcccccr}
\tablewidth{0pt}
\tabletypesize{\scriptsize}
\setlength{\tabcolsep}{0.01in}
\tablecaption{\textsl{GALEX} TDS and PS1 MDS\label{tdsmds}}
\tablehead{
\colhead{Survey} & \colhead{Field of View} & \colhead{Plate Xcale} & \colhead{PSF FWHM} &  \colhead{$m_{\rm lim}$ per Epoch} & \colhead{Cadence} & \colhead{Seasonal Visibility} \\
\colhead{} & \colhead{(deg)} &  \colhead{(arcsec/pixel)} & \colhead{(arcsec)}  & \colhead{(mag)} & \colhead{(days)} & \colhead{(months)}
}
\startdata
\textsl{GALEX} TDS & 1.1 & 1.5 & 5.3 & 23.3 & 2 & $\sim 1$ \\
PS1 MDS & 3.5 & 0.258 & 1.0 & 23.0 & 3 & $\sim 6 $ 
\enddata
\end{deluxetable}

\begin{deluxetable}{lrrcr}
\tablewidth{0pt}
\tabletypesize{\scriptsize}
\setlength{\tabcolsep}{0.01in}
\tablecaption{\textsl{GALEX} TDS Fields\label{tab1}}
\tablehead{
\colhead{Name} & \colhead{R.A. (J2000)} & \colhead{Dec (J2000)} & $E(B-V)$ & \colhead{Epochs} \\
\colhead{} & \colhead{(deg)} & \colhead{(deg)} & \colhead{(mag)} & \colhead{}
}
\startdata
PS\_XMMLSS\_MOS00 & 35.580 & $-$3.140 & 0.028 & 27\\
PS\_XMMLSS\_MOS01 & 36.500 & $-$3.490 & 0.025 & 27\\
PS\_XMMLSS\_MOS02 & 35.000 & $-$3.950 & 0.020 & 14\\
PS\_XMMLSS\_MOS03 & 35.875 & $-$4.250 & 0.026 & 27\\
PS\_XMMLSS\_MOS04 & 36.900 & $-$4.420 & 0.026 & 27\\
PS\_XMMLSS\_MOS05 & 35.200 & $-$5.050 & 0.022 & 26\\
PS\_XMMLSS\_MOS06 & 36.230 & $-$5.200 & 0.027 & 24\\
PS\_CDFS\_MOS00 & 53.100 & $-$27.800 & 0.008 & 114\\
PS\_CDFS\_MOS01 & 52.012 & $-$28.212 & 0.008 & 30\\
PS\_CDFS\_MOS02 & 53.124 & $-$26.802 & 0.009 & 29\\
PS\_CDFS\_MOS03 & 54.165 & $-$27.312 & 0.012 & 30\\
PS\_CDFS\_MOS04 & 52.910 & $-$28.800 & 0.009 & 30\\
PS\_CDFS\_MOS05 & 52.111 & $-$27.276 & 0.010 & 30\\
PS\_CDFS\_MOS06 & 53.970 & $-$28.334 & 0.010 & 6\\
PS\_COSMOS\_MOS21 & 150.500 &  $+$3.100 & 0.023 & 15\\
PS\_COSMOS\_MOS22 & 149.500 &  $+$3.100 & 0.027 & 16\\
PS\_COSMOS\_MOS23 & 151.000 &  $+$2.200 & 0.024 & 24\\
PS\_COSMOS\_MOS24 & 150.000 &  $+$2.200 & 0.020 & 26\\
PS\_COSMOS\_MOS25 & 149.000 &  $+$2.300 & 0.023 & 13\\
PS\_COSMOS\_MOS26 & 150.500 &  $+$1.300 & 0.023 & 26\\
PS\_COSMOS\_MOS27 & 149.500 &  $+$1.300 & 0.019 & 27\\
PS\_GROTH\_MOS01 & 215.600 &  $+$54.270 & 0.011 & 17\\
PS\_GROTH\_MOS02 & 213.780 &  $+$54.350 & 0.015 & 16\\
PS\_GROTH\_MOS03 & 214.146 &  $+$53.417 & 0.009 & 17\\
PS\_GROTH\_MOS04 & 212.400 &  $+$53.700 & 0.011 & 18\\
PS\_GROTH\_MOS05 & 215.500 &  $+$52.770 & 0.008 & 19\\
PS\_GROTH\_MOS06 & 214.300 &  $+$52.550 & 0.008 & 8\\
PS\_GROTH\_MOS07 & 212.630 &  $+$52.750 & 0.009 & 19\\
PS\_ELAISN1\_MOS10 & 242.510 &  $+$55.980 & 0.007 & 17\\
PS\_ELAISN1\_MOS11 & 244.570 &  $+$55.180 & 0.009 & 17\\
PS\_ELAISN1\_MOS12 & 242.900 &  $+$55.000 & 0.008 & 18\\
PS\_ELAISN1\_MOS13 & 241.300 &  $+$55.350 & 0.007 & 18\\
PS\_ELAISN1\_MOS14 & 243.960 &  $+$54.200 & 0.010 & 19\\
PS\_ELAISN1\_MOS15 & 242.400 &  $+$54.000 & 0.011 & 21\\
PS\_ELAISN1\_MOS16 & 241.380 &  $+$54.450 & 0.010 & 19\\
PS\_VVDS22H\_MOS00 & 333.700 &  $+$1.250 & 0.040 & 39\\
PS\_VVDS22H\_MOS01 & 332.700 &  $+$0.700 & 0.046 & 35\\
PS\_VVDS22H\_MOS02 & 334.428 &  $+$0.670 & 0.057 & 38\\
PS\_VVDS22H\_MOS03 & 333.600 &  $+$0.200 & 0.058 & 27\\
PS\_VVDS22H\_MOS04 & 334.610 &  $-$0.040 & 0.093 & 24\\
PS\_VVDS22H\_MOS05 & 333.900 &  $-$0.720 & 0.102 & 35\\
PS\_VVDS22H\_MOS06 & 332.900 &  $-$0.400 & 0.113 & 33
\enddata

\end{deluxetable}

\clearpage

\begin{deluxetable}{lrrrrrrrrrrrrr}
\tablewidth{6in}
\tabletypesize{\footnotesize}
\setlength{\tabcolsep}{0.01in}
\tablecaption{\textsl{GALEX} TDS Classifications \label{tab:class}}
\tablehead{
\colhead{} & \colhead{} & \multicolumn{2}{c}{Archive} & \multicolumn{2}{c}{PS1}  & \multicolumn{6}{c}{Classification}\\
\cline{8-12}\\
\colhead{Step} & \colhead{$N_{\rm unclass}$} & \colhead{pt} & \colhead{ext} & \colhead{pt} & \colhead{ext} & \colhead{orphan} & \colhead{QSO} & \colhead{RRL} & \colhead{Mdw} & \colhead{Star} & \colhead{AGN} & \colhead{pt} & \colhead{gal}
}
\startdata
optical match & 1078 & 487 & 391 & 76 & 103 & 21 &  & & & & & &\\
QSO color cut & 1078 &  326 & & & & & 326 &  &  & \\
RRL color cut & 753 &   37 & & & &  & & 37 &  &\\
Mdw color cut & 716 &  44 & & 9 & &  & & & 53 & \\
stellar locus color cut & 663 & 17 & & & & & & & & 17 \\
$S_{\rm yr}/S_{\rm d} \ge 3$ & 646 & 19 & & 15 & & & 34 &  \\
stochastic UV var  & 616 & & 200 & & 68  & & &  & &  & 268 \\
X-ray/spec match & 346 & 8 & 37 & & &  & 2 & & & 5 & 37 \\
\hline
 & 302 &  &  &  &  & 21 & 358 & 37 & 53 & 22 & 305 & 93 & 189 \\
\enddata
\end{deluxetable}

\begin{deluxetable}{lrrrrcccccccccc}
\tablewidth{0pt}
\tabletypesize{\footnotesize}
\setlength{\tabcolsep}{0.005in}
\tablecaption{\textsl{GALEX} TDS Catalog\tablenotemark{a} \label{tab2}}
\tablehead{
\colhead{} & \multicolumn{8}{c}{NUV} & \multicolumn{4}{c}{Optical} & \multicolumn{1}{c}{X-ray} & \multicolumn{1}{c}{Class} \\
\cline{10-12}  \\
\colhead{ID} & \colhead{R.A.} & \colhead{Dec} & \colhead{m$_{low}$} & \colhead{$|\Delta m_{\rm max}|$} & \colhead{$\sigma_{int}$} & \colhead{S$_{d}$} & \colhead{S$_{yr}$} & \colhead{LC} & \colhead{Morph} & \colhead{$r_{\rm AB}$} & \colhead{Color} & \colhead{$z$}  & \colhead{ } & \colhead{ }
}
\startdata
GROTH\_MOS01-21 & 216.1622 &  54.0911 & 22.54 & 4.60 & 1.04 & 0.80 & 1.13 & V & pt & 14.92 & Mdw &   &   & Mdw \\
VVDS22H\_MOS05-05 & 333.8326 &  -0.5491 & 21.14 & 4.47 & 0.99 & 0.90 & 0.74 & F & pt & 21.24 & QSO &   &   & CV \\
ELAISN1\_MOS15-02 & 242.0397 &  54.3586 & $>$22.89 & $>$4.03 & 1.37 & 0.84 & 2.93 & V & pt & 22.51 & QSO &   &   & CV \\
ELAISN1\_MOS15-09 & 242.3685 &  53.6738 & $>$23.03 & $>$4.02 & 1.23 & 0.01 & 2.15 & F & ext & 21.05 &   &   &   & Galaxy Trans \\
GROTH\_MOS07-09 & 212.5024 &  52.4153 & $>$23.10 & $>$3.61 & 3.51 & 0.09 & 0.69 & F & pt & 17.45 & Mdw &   &   & Mdw \\
CDFS\_MOS02-20 &  53.1682 & -26.3564 & $>$23.06 & $>$3.60 & 0.73 & 0.95 & 0.78 &   & \nodata & 15.72 & Mdw &   &   & Mdw \\
CDFS\_MOS00-41 &  53.3453 & -27.3361 & $>$23.10 & $>$3.36 & 0.36 & 0.39 & 0.15 & V & \nodata & 15.76 &   &   &   & Mdw \\
COSMOS\_MOS22-11 & 149.4973 &   3.1171 & 22.56 & 3.35 & 0.75 & \nodata & 0.58 & F & pt & 14.25 & Mdw &   &   & Mdw \\
GROTH\_MOS05-00 & 214.9435 &  52.9953 & 22.07 & 3.29 & 0.71 & 1.18 & 0.65 & F & pt & 14.15 & Mdw &   & X & Mdw \\
XMMLSS\_MOS06-22 &  36.6468 &  -5.0886 & $>$22.95 & $>$3.28 & 0.95 & $<$0 & 4.04 &   & pt & 15.58 & Mdw &   & X & Mdw
\enddata
\tablenotetext{a}{Table 4 is published in its entirety in the electronic edition of ApJ.  A portion is shown here for guidance regarding its format and content.}
\end{deluxetable}

\end{document}